\documentclass[reprint,aps,prd,groupedaddress,nofootinbib,superscriptaddress,eqsecnum,amsmath,amssymb]{revtex4}
\usepackage{amsfonts}
\usepackage{graphicx}
\usepackage{bm}
\usepackage{float}
\usepackage[colorlinks=true,
    linkcolor=blue,
    citecolor=blue,
    urlcolor=blue]{hyperref}
  \usepackage{tikz}
\usetikzlibrary{arrows.meta}
\usetikzlibrary{shapes.geometric,arrows}   
\usepackage[figurename=Fig.]{caption}

\newcommand{\bse}[1]{\begin{subequations}\begin{eqnarray} #1 \end{eqnarray}\end{subequations} }
\newcommand{\ernst}{\mathcal{E}}
\renewcommand\Re{\operatorname{\mathsf{Re}}}
\renewcommand\Im{\operatorname{\mathsf{Im}}}

\newcommand{\diff}{\mathrm{d}}
\newcommand{\bp}[1]{\left( #1 \right)}

\newcommand{\bb}[1]{\left[ #1 \right]}
\newcommand{\bbr}[1]{\left\lbrace #1 \right\rbrace}
\newcommand{\expo}[1]{\mathrm{e}^{#1}}
\newcommand{\nqform}[1]{\boldsymbol{\nabla}#1\cdot\boldsymbol{\nabla}#1}
\newcommand{\qform}[1]{#1\cdot#1}
\newcommand{\bv}[1]{\hat{\boldsymbol{#1}}}
\newcommand{\bn}{\boldsymbol{\nabla}}

\newcommand{\sfE}{\mathsf{E}}
\newcommand{\sfH}{\mathsf{H}}
\newcommand{\sfT}{\mathsf{T}}

\newcommand{\oo}{\hat{\omega}}

\begin{document}



\title{Pleban\'ski--Demia\'nski \`a la Ehlers--Harrison: Exact Rotating and Accelerating Type I Black Holes}

\author{Jos\'e Barrientos}
\email{jbarrientos@academicos.uta.cl}
\affiliation{Sede Esmeralda, Universidad de Tarapac\'a, Avenida Luis Emilio Recabarren 2477, Iquique, Chile}
\affiliation{Institute of Mathematics of the Czech Academy of Sciences, \v{Z}itn\'a 25, 115 67 Praha 1, Czech Republic}

\author{Adolfo Cisterna}
\email{adolfo.cisterna@mff.cuni.cz}
\affiliation{Sede Esmeralda, Universidad de Tarapac\'a, Avenida Luis Emilio Recabarren 2477, Iquique, Chile}
\affiliation{Institute of Theoretical Physics, Faculty of Mathematics and Physics,
Charles University, V Hole{\v s}ovi{\v c}k{\' a}ch 2, 180 00 Praha 8, Czech Republic}

\author{Konstantinos Pallikaris}
\email{konstantinos.pallikaris@ut.ee}
\affiliation{Laboratory of Theoretical Physics, Institute of Physics, University of Tartu, W. Ostwaldi 1, 50411 Tartu, Estonia}

\begin{abstract}
Recently, it was shown that type D black holes, encompassed in the large Pleban\'ski--Demia\'nski (PD) family, exhibit a wide class of algebraically general generalizations via the application of Ehlers and Harrison transformations. In this work, we first discuss some mathematical details behind the composition of such transformations, and next, we introduce a qualitative picture of the most general type I generalization of the PD family, dubbed ``Enhanced Pleban\'ski--Demia\'nski'' spacetime. We provide the exact form of the solution in the original PD coordinates, obtained via the simultaneous action of an Ehlers and a Harrison transformation on the vacuum PD geometry. In order to make the physics more transparent, we explicitly construct a rotating and accelerating black hole which further has NUT parameter and electric charges, both of them entering, not only the event horizon, but the Rindler horizon as well. This solution is directly obtained in the ``physical'' coordinates recently proposed by Podolsk\'y and Vr\'atny. Finally, a pedagogical appendix is thoughtfully included, providing readers with a user-friendly step-by-step guide to the Ernst formalism, in an attempt to address and resolve various minor inconsistencies frequently appearing in the relevant literature.

\end{abstract}

\maketitle

\section{Introduction}

The Kerr black hole is a particularly noteworthy exact solution of Einstein's field equations, especially from an astrophysical point of view. Realistic celestial bodies generally exhibit rotational motion. Even if their rotation is minimal, the conservation of angular momentum turns out to play an important role during gravitational collapse. Consequently, an analytical expression for the exterior spacetime around rotating sources is mandatory for studying such scenarios, a fact (among others) showcasing the overall significance of studying exact solutions within the framework of General Relativity (GR).

{ The Pleban\'ski--Demia\'nski (PD) family of solutions \cite{Plebanski:1976gy,Podolsky:2021zwr,Podolsky:2022xxd} plays a crucial role in understanding the Einstein–Maxwell field equations. This family is considered the most general solution of type D spacetimes, a classification based on the Petrov classification scheme \cite{Stephani:2003tm}. The PD family encompasses a wide range of solutions, including well-known black hole solutions like the Kerr and Reissner--Nordstr\"om (RN) metrics. Its causal structure is particularly interesting as it models scenarios involving two rotating and charged black holes that are accelerating away from each other  \cite{Bicak:1999sa,Pravda:2002kj,Griffiths:2005se,Griffiths:2005mi,Griffiths:2005qp}.
}

A distinctive observation has been made concerning the PD hierarchy of solutions, revealing the absence of a nonrotating limit when both acceleration and NUT charge are present \cite{Griffiths:2009dfa}. This led to the conjecture that accelerating NUT black holes may not exist, or, if they do exist, that they do not belong to the PD class, suggesting that they might not be found among algebraically special spacetimes. 
Despite the inherent challenges in understanding accelerating NUT spacetimes, Chang, Mann, and Stelea managed to construct a sort of accelerating NUT black hole in their seminal work~\cite{Chng:2006gh}, employing intricate solution-generating techniques. Based on the SL(2,$\mathbb{R}$) symmetry of a reduced Lagrangian obtained via dimensional reduction of the four-dimensional GR Lagrangian along the time direction, they successfully demonstrated that, with an accelerating version of the Zipoy--Voorhees line element~\cite{Griffiths:2009dfa} as a seed, one can obtain a new solution which correctly reduces to the Taub--NUT black hole in the zero-acceleration limit, while it also assumes the standard form of the C-metric in a certain parameter limit.
{Thorough analysis of this solution was later conducted by Podolsk\'y and Vr\'atny~\cite{Podolsky:2020xkf},  showing that it represents a genuine accelerating NUT black hole, and, moreover, that it falls under algebraic type I, thereby being algebraically general. Consequently, the aforementioned solution is not included in the PD family. 
Remarkably, the NUT parameter not only enters the black hole horizons but the accelerating horizons as well, as it can be directly observed from the structure of the lapse function \cite{Podolsky:2020xkf,Barrientos:2023tqb,Astorino:2023elf}.} 

Recently, a highly efficient mechanism for introducing NUT charge to accelerating spacetimes has been proposed~\cite{Barrientos:2023tqb,Astorino:2023elf}. This innovative approach is based on the utilization of Ehlers transformations \cite{Ehlers:1957zz, Ehlers:1959aug,  Harr}, part of the Lie point symmetries inherent in the Einstein--Maxwell system, which become apparent when expressing the action in terms of the Ernst potentials~\cite{Ernst:1967wx,Ernst:1967by}. Through the application of the so-called electric Ehlers transformation, the proposed method adds a NUT charge to a given seed spacetime. In particular, it allows the introduction of a single NUT charge \cite{Barrientos:2023tqb}, or even two such charges~\cite{Astorino:2023elf}, to any stationary axially symmetric spacetime in electrovacuum.

In \cite{Barrientos:2023tqb}, it has been demonstrated how the above machinery accurately provides the Chang, Mann, and Stelea solution~\cite{Chng:2006gh} with remarkable simplicity, following the approach proposed by Podolsk\'y and Vr\'atny \cite{Podolsky:2020xkf}. Additionally, a Reissner--Nordstr\"om C-metric NUT black hole that faithfully reduces to the RN-C-metric and RN-NUT configurations in certain limits, has been also presented.\footnote{Extension of these solutions by including a conformally coupled scalar in the matter sector, has also been studied in \cite{Barrientos:2023tqb}.} While electric Ehlers transformations were readily known to add NUT charge to a given seed \cite{Vigano:2022hrg}, the primary focus was directed towards static spherically symmetric seeds. As a result, the intricate interplay between NUT charge and the accelerating nature of a given seed has not been given much attention. The principal novelty of considering accelerating seeds lies in the emergence of Rindler horizons, representing the causal obstructions experienced by any accelerating observer along her/his trajectory. An Ehlers transformation does not only affect the black hole horizons but also the Rindler horizons, yielding novel backreactions involving a NUT-enhanced background. 

A valuable understanding of the altered Rindler horizons that these type I accelerating black holes exhibit can be gained by considering these solutions as given by taking a specific limit on the metric of black hole binaries {\cite{Astorino:2023ifg, Astorino:2023uim}}. It's worth recalling that the near-horizon structure of a Schwarzschild black hole is delineated by the Rindler metric, characteristic of an accelerating observer. {A mathematically valid approach to focus on the near horizon geometry is to actually perform the infinite mass limit of the solution. Visualizing an accelerating Schwarzschild black hole as a binary system of two Schwarzschild black holes, effectively described by the Bach–Weyl solution {\cite{BachWeyl, ISRAEL1964}}, involves one of the two black holes growing infinitely large (achieving infinite mass) while maintaining a finite distance from the other. The event horizon of the ``big'' black hole then manifests as an accelerating horizon to its smaller counterpart. Consequently, the Bach-Weyl solution takes on the appearance of a C-metric in this limit, as was shown by Wang \cite{Wang:1996sn}.} Similarly, these type I accelerating black holes, featuring Rindler horizons dependent on transformation parameters, among other factors, can be envisioned as a limit of the NUTty and/or charged extension of the Bach-Weyl spacetime. This argument support what was naively understood from the structure of the Rindler horizons of the accelerating-NUT black hole described \cite{Barrientos:2023tqb,Astorino:2023elf}, where it can be seen via the different changes of coordinates how the geometry is centered along the inner horizon $r_-=m-\sqrt{m^2+l^2}$, therefore how the NUT parameter pervades the Rindler horizon.

The investigation of incorporating a second NUT parameter, into a solution that already carries NUT charge, has been explored in~\cite{Astorino:2023elf}. As expected, in the case of nonaccelerating seeds, the introduction of a second NUT parameter proves to be redundant, as the latter can be absorbed by the NUT charge already present in the spacetime. However, for accelerating seeds with angular momentum, a distinct scenario unfolds; both NUT charges, the one confined to the horizon, and the other permeating throughout the whole spacetime, can in general coexist. Tuning both NUT charges, doable only in the presence of angular momentum, proves to be useful for removing the Misner string. Such a scenario has been studied considering the full PD class.

The discovery of these findings has sparked a renewed interest in probing the black hole spectrum of GR beyond the well-explored type D class, leading to novel ways for constructing algebraically general black hole solutions. Among the evident extensions to be considered, lies the application of electric Harrison transformations~\cite{Harr}, which are known to introduce electric and magnetic monopolic charges to a given seed. Indeed, when applying a Harrison transformation to add electric charge to a C-metric seed, it becomes apparent that the resulting solution is not algebraically special, but of algebraically general nature instead. Once again, a key element in this construction revolves around the occurrence of Rindler horizons. Both event and Rindler horizons are imbued with electric charge, a fact strongly affecting the spacetime geometry. The newly obtained charged accelerating solution is different from the well-known RN-C-metric present in the PD family. This indicates that the addition of electric charge to an accelerating seed leads to a unique class of type I black holes with distinctive properties. The construction of such type I charged accelerating black holes has been initially addressed in~\cite{Astorino:2023ifg}, with a particular emphasis given on the RN-C-metric and RN-C-metric-NUT cases.

In this study, we present a qualitative picture of the most general 
type I extension of the PD family, achievable by the sequential application of Ehlers and Harrison transformations, a direct application of their composition that is. The resulting spacetime does in general feature two distinct NUT parameters and two sets of electromagnetic charges. We term this final configuration the ``Enhanced Pleban\'ski-Demia\'nski'' spacetime (EPD). In the physical spherical-like coordinates introduced by Podolsk\'y and Vr\'atny \cite{Podolsky:2021zwr,Podolsky:2022xxd}, its most general form would be described by nine parameters, the six parameters contained in the original PD spacetime, i.e., the mass $m$, { Kerr-like rotation parameter $a$}, acceleration parameter $A$, NUT charge $l$ and electromagnetic charges $e$ and $g$, together with a second NUT parameter $\bar{l}$ and a second pair of electromagnetic charges $\bar{e}$ and $\bar{g}$, induced by the Ehlers and Harrison maps, respectively. The latter three are henceforth dubbed Ehlers--Harrison charges. {Notice that $\bar{l}$, $\bar{e}$ and $\bar{g}$ will correspond to a reparametrization of the original parameters introduced via the Ehlers--Harrison map. Due to the high computational complexity of the task, we are able to provide the explicit form of the EPD spacetime in the original PD coordinates, only for a neutral PD seed. Despite this, the solution we present is sufficiently general and novel. Although the use of the original PD coordinates (and parameters) proves to be mandatory for the integration of the solution, regarded as a computational problem per se, the physical meaning is more or less obscure. For this reason, we also explicitly provide the full spacetime of an accelerating and rotating black hole carrying Ehlers--Harrison charges, using the physics-wise transparent form of the PD metric~\cite{Podolsky:2021zwr,Podolsky:2022xxd} as the seed.}

Our paper is structured as follows: in Sec. \ref{secErnst}, we provide a concise introduction to Ehlers and Harrison transformations and discuss their crucial role in generating novel stationary axially symmetric solutions within the Einstein--Maxwell framework. Furthermore, we thoroughly investigate compositions of these transformations, disclosing an interesting equivalence (under certain assumptions) between the composition of two Harrison transformations and that of an Ehlers transformation with a Harrison one. In Sec. \ref{secEPD}, we present the Enhanced Pleban\'ski--Demia\'nski type I hierarchy of solutions and explicitly construct the EPD spacetime in PD coordinates, starting from a neutral PD seed. Next, we provide (in spherical-like coordinates) an exact expression for the metric representing an accelerating and rotating black hole with both Ehlers and Harrison charges, together with an expression for the gauge field supporting it. Various limits are discussed. We conclude our study in Sec. \ref{seccoments}, where we highlight the significance of the new findings and discuss promising ways for further exploration using these innovative techniques. Lastly, in Appendix \ref{App2}, we offer a user-friendly rederivation of the Ernst equations, in an attempt to address sign inconsistencies often appearing in the relevant literature.

\section{Ernst equations and the SU(2,1) symmetry}\label{secErnst}

The mathematical framework developed by Ernst in the 1960s \cite{Ernst:1967wx,Ernst:1967by} has been a particularly valuable tool for studying stationary axisymmetric electrovacuum fields. Its remarkable novelty is the disclosure of additional symmetries in the Einstein--Maxwell system which remain elusive in the standard formulation. By casting the Einstein--Maxwell field equations into a set of two complex equations for the complex Ernst potentials, one ends up finding a collection of symmetry transformations which form a Lie group with eight real parameters~\cite{kinnersley1973generation,Kinnersley:1977pg}, isomorphic to SU(2,1). 

In a nutshell, the formulation works as follows.\footnote{See Appendix \ref{App2} for a detailed derivation of the Ernst equations. Here, we are assuming the so-called ``electric'' version of the LWP spacetime, and we have set $p=0$ and $s=1$.} The most general stationary and axially symmetric spacetime within the Einstein--Maxwell framework is represented by the well-known Lewis--Weyl--Papapetrou (LWP) line element and the gauge field accompanying it, 
\bse{
\diff s{}^2 &=& -f\bp{\diff t-\omega \,\diff\varphi}^2 + \frac{1}{f}\bb{\rho^2\,\diff\varphi^2 + \expo{2\gamma}\bp{\diff \rho^2 + \diff z^2}},\label{eq:e-LWPMain}\\
A&=&A_t\,\diff t+A_\varphi\,\diff\varphi,
}
respectively, where $f,\ \omega$, and $\gamma$ are functions of Weyl's coordinates $\rho$ and $z$. It can be shown (see Appendix~\ref{App2} for details) that, defining the pair of (complex) Ernst potentials
\begin{equation}
    \ernst = f-|\Phi|^2+i\chi,\quad \Phi = A_t+i\Tilde{A}_\varphi,\label{eq:ErnstPotentials}
\end{equation}
the Einstein--Maxwell field equations are cast into two complex three-dimensional equations, namely
\begin{subequations}\label{eq:ErnstEquationsMain}
    \begin{eqnarray}
        \bp{\Re\ernst + |\Phi|^2}\nabla^2\ernst &=& \bn\ernst\cdot\bp{\bn\ernst + 2\Phi^*\bn\Phi},\\
        \bp{\Re\ernst + |\Phi|^2}\nabla^2\Phi &=& \bn\Phi\cdot\bp{\bn\ernst + 2\Phi^*\bn\Phi}.
    \end{eqnarray}
\end{subequations}
Here, all vector quantities are understood as vectors in flat space with cylindrical coordinates $\{\rho,z,\varphi\}$. The so-called twisted potentials $\Tilde{A}_\varphi$ and $\chi$ are then given by
\begin{subequations}\label{eq:TwistEquations}
\begin{eqnarray}
    \hat{\boldsymbol{\varphi}}\times \boldsymbol{\nabla}\Tilde{A}_\varphi &=& \frac{ f}{\rho}\bp{\bn A_\varphi + \omega \bn A_t},\\
    \bv{\varphi}\times\bn\chi &=&-\bp{\frac{f^2}{\rho}\bn\omega + 2\bv{\varphi}\times\Im\bp{\Phi^*\bn\Phi}}, 
\end{eqnarray}
\end{subequations}
respectively.

{Equations~\eqref{eq:ErnstEquationsMain} enjoy certain symmetries, which must then be inherent in the Einstein-Maxwell system. These symmetry transformations, which are henceforth referred to as Ernst symmetries, are
\begin{subequations}
\label{ernst-group}
\begin{align}
\label{gauge1}
\mathsf{T}^1_b:\quad\ernst & = \ernst_0 + ib \,, \qquad\qquad\qquad\quad\,
\Phi = \Phi_0 \,, \\
\label{gauge2}
\mathsf{T}^2_\alpha:\quad\ernst & = \ernst_0 - 2\alpha^*\Phi_0 - |\alpha|^2 \,, \quad\quad\hspace{0.1cm}
\Phi = \Phi_0 + \alpha \,, \\
\label{gauge3}
\mathsf{T}^3_\lambda:\quad\ernst & = |\lambda|^2 \ernst_0 \,, \qquad\qquad\qquad\quad\;\;
\Phi = \lambda \Phi_0 \,, \\
\label{ehlers}
\mathsf{E}_c:\quad\ernst & = \frac{\ernst_0}{1 + ic\ernst_0} \,, \qquad\qquad\qquad\;\,
\Phi = \frac{\Phi_0}{1 + ic\ernst_0} \,,  \\
\label{harrison}
\mathsf{H}_\beta:\quad\ernst & = \frac{\ernst_0}{1 - 2\beta^*\Phi_0 - |\beta|^2\ernst_0} \,, \quad \hspace{0.2cm}
\Phi = \frac{\beta\ernst_0 + \Phi_0}{1 - 2\beta^*\Phi_0 - |\beta|^2\ernst_0} \,, 
\end{align}
\end{subequations}
where $\alpha, \beta$ and $\lambda$ are complex parameters, while $b$ and $c$ are real.\footnote{In this section, latin letters $a,b,\ldots$ are reserved for real parameters, whereas Greek letters $\alpha,\beta,\ldots$ stand for complex ones.} Not all of these transformations can be used to generate novel spacetimes. In fact, \eqref{gauge1} and \eqref{gauge2} are nothing else than gravitational and electromagnetic gauge transformations, while \eqref{gauge3}
corresponds to a coordinate rescaling combined with an electromagnetic duality rotation. 
However, the remaining symmetries, \eqref{ehlers} and \eqref{harrison}, the so-called Ehlers \cite{Ehlers:1957zz} and Harrison \cite{Harr} transformations, act in a nontrivial way, thereby producing new nonequivalent spacetimes which, of course, are again solutions of the Einstein--Maxwell field equations. 

Additionally, the Ernst equations  also possess a discrete ``inversion'' transformation, namely
\begin{equation}
    \mathbf{I}: \bp{\ernst_0,\Phi_0}\mapsto \bp{\frac{1}{\ernst_0},\frac{\Phi_0}{\ernst_0}}.
\end{equation}
With these at hand, it is quite straightforward to observe that a certain composition of the above transformations leads to the Ehlers and Harrison transformations \cite{Stephani:2003tm}.}

In the next subsection, we focus on compositions of Ehlers and Harrison transformations, which will later be used in Sec.~\ref{secEPD} to construct new stationary and axially symmetric solutions of the Einstein--Maxwell system, which are algebraically general. 

\subsection{Compositions}

Due to the fact that the composition of two inverse transformations equals the identity transformation, i.e., $\mathbf{I}\circ\mathbf{I}=\mathbb{I}$, and since gravitational gauge transformations commute, namely $\mathsf{T}^1_{b}\circ \mathsf{T}^1_{c}=\mathsf{T}^1_{b+c}$, it easily follows that Ehlers transformations form a one-parameter subgroup; they satisfy the group property $\mathsf{E}_{b}\circ\mathsf{E}_{c}=\mathsf{E}_{b+c}$. Moreover, since, in general, $\mathsf{T}^1_c\circ \mathsf{T}^2_\beta = \mathsf{T}^2_\beta\circ \mathsf{T}^1_c$, it also follows that Ehlers transformations commute with Harrison ones, viz., $\mathsf{E}_{c}\circ\mathsf{H}_{\beta} =\mathsf{H}_{\beta} \circ \mathsf{E}_{c}$. Actually, since in the next section we are going to use this particular composition to build the new solutions, let us be proactive and display this map here, 
\begin{eqnarray}
         \mathsf{E}_{c}\circ\mathsf{H}_{\beta}:\bp{\ernst_0,\Phi_0}\mapsto\bp{\frac{\ernst_0}{1- 2\beta^*\Phi_0+\bp{ic - |\beta|^2} \ernst_0 }, \frac{\beta \ernst_0 + \Phi_0}{1- 2\beta^*\Phi_0+\bp{ic - |\beta|^2} \ernst_0 }}.\label{eq:EHpotentials}
\end{eqnarray}

On the other hand, we observe that Harrison transformations fail to form a subgroup, this due to the fact that the electromagnetic gauge transformations do not commute,
\begin{equation}
    \bp{\mathsf{T}^2_{\beta}\circ\mathsf{T}^2_{\alpha} - \mathsf{T}^2_{\alpha+\beta}}\bp{\ernst_0,\Phi_0} = \beta \alpha^* - \alpha\beta^*.
\end{equation}
Indeed, $\mathsf{T}^2_{\beta}\circ\mathsf{T}^2_{\alpha} = \mathsf{T}^1_{i\bp{\alpha\beta^* - \beta\alpha^*}}\mathsf{T}^2_{\alpha+\beta}$, which implies that two general Harrison transformations amount to a particular Ehlers--Harrison one, namely, 
\begin{equation}
    \mathsf{H}_{\alpha}\circ\mathsf{H}_{\beta} = \mathsf{E}_{i\bp{\alpha\beta^* - \beta\alpha^*}}\mathsf{H}_{\alpha+\beta}.\label{eq:HHcomp}
\end{equation}
This is quite an interesting observation. Recall that the Harrison map is thought of as a charging transformation rendering vacuum into electrovacuum solutions. But what does really happen when the seed is an electrovacuum one?\footnote{Remember that an electrovacuum solution can always be obtained via a Harrison transformation of a vacuum seed.} The above composition property seems to tell us that the application of a Harrison transformation on a static electrovacuum seed will lead to a stationary electrovacuum spacetime---albeit suffering from a NUT-like singularity. Of course, this NUT parameter will not be free; it is rather determined by the other parameters and charges at play. 

All of this is well understood in the light of Eq.~\eqref{eq:HHcomp}. Two general Harrison transformations applied to a vacuum seed amount to charging this seed and simultaneously adding a {fixed} NUT parameter to it. Therefore, without resorting to further compositions of the Harrison map with the other symmetries, the only way to avoid the cross term in the target metric is if $\alpha,\beta$ satisfy the relation
\begin{equation}
    \Re\beta= \frac{\Re\alpha}{\Im\alpha}\Im\beta,
\end{equation}
in which case $\sfH_\beta\circ\sfH_\alpha = \sfH_{\alpha+\beta}$. For example, assuming $\Im\alpha\neq 0$, we have that
\begin{equation}
    \mathsf{H}_{c(\Re\alpha/\Im\alpha +i)}\circ \mathsf{H}_{\alpha}=\mathsf{H}_{\alpha(1+c/\Im\alpha)},
\end{equation}
where we remind the reader that $\alpha$ is complex whereas $c$ is real. The generalization is straightforward; since a composition of Ehlers transformations is again an Ehlers transformation, and since Ehlers transformations commute with Harrison ones, from Eq.~\eqref{eq:HHcomp} we can conclude that 
\begin{equation}
    \mathsf{H}_{\alpha_p}\circ\ldots\circ \mathsf{H}_{\alpha_2}\circ\mathsf{H}_{\alpha_1} = \mathsf{E}_c\circ \mathsf{H}_{\alpha_1+\alpha_2+\ldots+\alpha_p},
\end{equation}
where the real parameter $c$ is fixed in terms of the parts of $\alpha_1,\alpha_2,\ldots,\alpha_p$.

As we are going through the various composition properties, and since the enhanced transformations presented in~\cite{Astorino:2023ifg} and~\cite{Astorino:2019ljy} prove to be convenient, it is worth studying the latter in the above spirit. The so-called enhanced Ehlers transformation
\begin{equation}
    \mathsf{EE}_c:\bp{\ernst_0,\Phi_0}\mapsto \bp{\frac{\ernst_0+ic}{1+i c\ernst_0},\Phi_0\frac{1+i c}{1+i c\ernst_0}},
\end{equation}
which directly provides purely the NUT extension of a given seed, is nothing else than the composition
\begin{equation}
    \mathsf{EE}_c = \mathsf{T}^1_c\circ\mathsf{T}^3_{1+ic}\circ \mathsf{E}_c.
\end{equation}
Enhanced Ehlers transformations retain the properties of the original Ehlers transformations in the sense that they also form a one-parameter subgroup,  
\begin{equation}
    \mathsf{EE}_b\circ\mathsf{EE}_c = \mathsf{EE}_{(b+c)/(1-bc)}.
\end{equation}
It is also fortunate that the enhancing itself is an operation which can be applied after transforming the solution \`a la Ehlers. However, this is not the case for the enhanced version of the Harrison transformation presented in~\cite{Astorino:2023ifg}, which reads 
\begin{equation}
    \mathsf{EH}_{\{\alpha,b\}} = \sfT^2_{-c\alpha}\circ \sfH_\alpha\circ \sfT^3_{c\expo{ib}},\quad c=\frac{\sqrt{1+4|\alpha|^2}-1}{2|\alpha|^2},
\end{equation}
for the Harrison transformation does not commute with $\mathsf{T}^3$.\footnote{Note here that Ehlers and Harrison transformations do not in general commute with the gauge transformations.} The enhanced version introduces one additional real parameter $b$, besides the Harrison transformation parameter $\alpha$. The former can be appropriately fixed to nullify the cross-term contribution from the Harrison operation in the target metric, if seed charges are present.

\section{Enhanced Pleban\'ski--Demia\'nski metric: The type I hierarchy}\label{secEPD}

\begin{figure}[h]
\begin{center}
    \includegraphics[width=\textwidth]{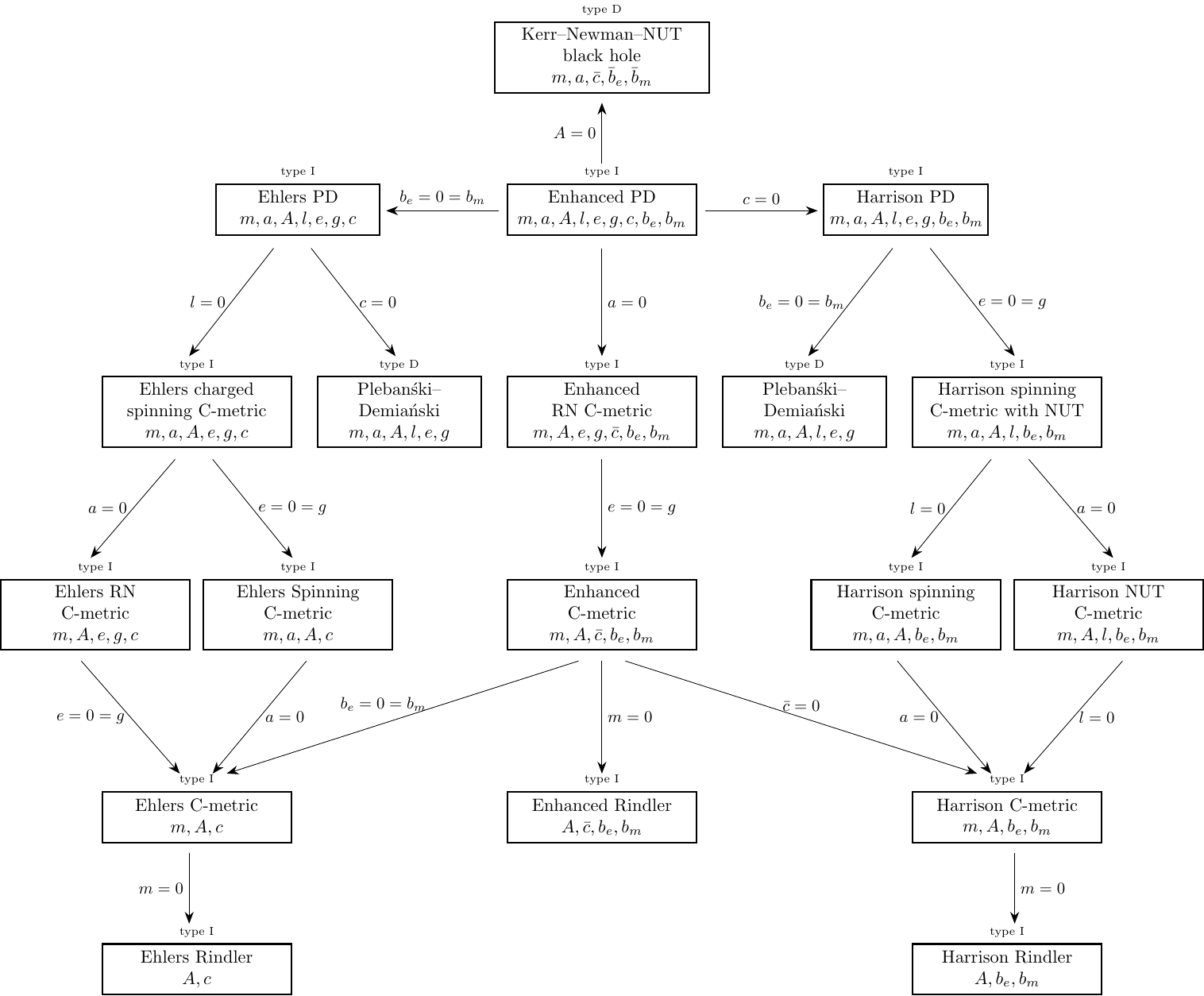}
\end{center}
\caption{Hierarchy of solutions for the Enhanced-Pleban\'ski--Demia\'nski spacetime.}\label{FIG1}
\end{figure}

Having delineated the transformations in Sec.~\ref{secErnst} and armed with a clear understanding of the effect that the Ehlers--Harrison transformation has on the accelerating horizons, we are well-positioned to seek the explicit integration of the all-inclusive family of type I geometries, herein referred to as the Enhanced Pleban\'ski--Demia\'nski spacetime, or EPD for short. As the task of integrating Eqs.~\eqref{eq:TwistEquations} for the twisted potentials in this most general case, proves to be a rather daunting computational challenge in spherical-like coordinates, we shall use a ``lighter'' form of the PD metric, the one in the original PD coordinates. We banish the details to Appendix~\ref{App1}, with the immensity of the expressions justifying us in doing so. Based on the findings therein, we are well-equipped to provide a description of the entire hierarchy tree of solutions within this new EPD family, which features in~Fig.~\ref{FIG1} as the ``parent'' spacetime. Although we lack an analytic form for the full EPD spacetime, the one with all seed parameters and charges switched on, we nevertheless corroborate the tree structure in~Fig.~\ref{FIG1} with our results in Appendix~\ref{App1}, at least up to a case directly next to the most general one, in particular, the EPD solution without seed electromagnetic charges.

{It is worth noting that, in the hierarchy diagram, the root node (the EPD spacetime) is characterized by a set of physical parameters $\{m,a,A,l,e,g\}$ of the PD seed and by an additional parameter triplet introduced via the Ehlers--Harrison operation and denoted as $\{c,b_e,b_m\}$ hereafter. The parameters $b_e$ and $b_m$ are the real and imaginary parts, respectively, of a Harrison parameter $\beta$, whereas $c$ stands for a real Ehlers transformation parameter. Recall that $b_e$ is then associated with the inclusion of electric charge, while $b_m$ with the inclusion of a magnetic monopolic charge. 
Since the results in Appendix~\ref{App1} have been derived using the original PD coordinates, it is worth remarking that establishing the relationships between the parameters in the \`a la PD form of the target metric and the physical parameters we actually use in the hierarchy structure, can be tricky at times, yet a definitely feasible task. Therefore, the use of the physical parameters in Fig.~\ref{FIG1} is \emph{ipso facto} justified.

Moreover, the explicit parametrizations of the Ehlers--Harrison parameters $\{c,b_e,b_m\}$ in terms of the extra NUT and electromagnetic charges $\{\bar{l},\bar{e},\bar{g}\}$, and {vice versa}, which necessarily contain (some of) the seed parameters, may vary between the various nodes in the hierarchy tree. Thus, we prefer to adhere to the use of $\{c,b_e,b_m\}$ in most cases. In general, a zoo of reparametrizations is usually necessary to present the various metrics in the standard form, or in a desired form in lack of a standard one. Having said that, the unfortunate occurrence of the same symbols in various ``children'' of the EPD spacetime in Fig.~\ref{FIG1}, should not mislead the reader into believing that the parameters are actually the same (although such cases are not excluded). They should rather be understood in terms of the ``physical'' properties they characterize, these being the same in all cases included. Then, finding the specific reparametrizations, is a work better undertaken on a case-by-case basis.}

We shall also remark that the term ``enhanced'' is used to convey the action of an Ehlers--Harrison map, thereby effecting a nontrivial transformation of the background, in which the original PD spacetime resides. Here, all the extra parameters, that is, $c$, $b_e$, and $b_m$, appear. When we only operate with one of the two maps, we refer to the resulting family as either ``Ehlers'', or ``Harrison'', with $c$ entering the solution in the former case, and $b_e$ and $b_m$ in the latter. For example, Ehlers-RN-C-metric corresponds to an RN-C-metric black hole in a background with NUT parameter $c$, brought in via the Ehlers transformation. Similarly, Harrison PD indicates a PD spacetime in a background featuring electromagnetic charges $b_e$ and $b_m$, added via the Harrison operation.

At this stage, it is also important to mention two key cases. {First, in the vanishing-acceleration limit, keeping two sets of NUT and electromagnetic charges (seed and Ehlers--Harrison), is redundant. After an appropriate reparametrization, only one set of charges should remain. Second, exclusively in the presence of rotation, that is $a\neq0$, the seed NUT parameter $l$ can coexist with the Ehlers NUT. When $a=0$, the presence of two NUT parameters is again superfluous, and only a single effective NUT parameter should remain. On the other hand, the seed electromagnetic charges $e$ and $g$ can exist as independent charges, along the Harrison charges, even for a nonrotating seed.} Consequently, in an attempt to have a consistent notation for all cases depicted in Fig.~\ref{FIG1}, we lastly adhere to the following rule. For effective NUT and electromagnetic charges, i.e., combinations of seed NUT with Ehlers NUT and combinations of seed charges with Harrison charges, respectively, we use a bar accent. Thus, since we previously agreed to use of $c,b_e,b_m$ in the hierarchy tree, we shall also use $\bar{c},\,\bar{b}_e$, and $\bar{b}_m$ to denote the effective quantities.

Finally, when classifying the solutions, a convenient criterion to distinguish if a given spacetime is algebraically general or special, is to examine the relation 
\begin{equation}\label{identity}
I^3=27 J^2,
\end{equation}
where
\begin{equation}
I=\Psi_0\Psi_4-4\Psi_1\Psi_3+3\Psi_2^2,\qquad J=\begin{vmatrix}\Psi_{0} & \Psi_{1} & \Psi_{2}\\
\Psi_{1} & \Psi_{2} & \Psi_{3}\\
\Psi_{2} & \Psi_{3} & \Psi_{4}
\end{vmatrix}.
\end{equation}
A spacetime is said to be algebraically general, ergo of Petrov type I, whenever the identity~\eqref{identity} is not satisfied. Otherwise, the spacetime is said to be algebraically special. Then, a convenient strategy to follow here, is to choose a tetrad, for which, the invariants $\Psi_1$ and $\Psi_3$ vanish, with $\Psi_0,\Psi_4\neq 0$. This allows one to write Eq.~\eqref{identity} in the simpler form
\begin{equation}
\Psi_0\Psi_4\left(\Psi_0\Psi_4-9\Psi_2^2\right)^2=0,
\end{equation}
which implies that, if $\Psi_0\Psi_4\neq9\Psi_2^2$, the spacetime is algebraically general.

In the subsequent subsection, we construct an accelerating and rotating black hole endowed with NUT and electromagnetic charges entering both horizons, Rindler and black hole ones. We attain this solution by acting with the Ehlers--Harrison map on a neutral NUTless PD seed. This time, we derive the solution in the physical spherical-like coordinates \emph{ab initio}, a fact compensating for the sacrifice of yet another seed parameter. 

\subsection{Enhanced Kerr: accelerating and rotating black hole with NUT parameter and electromagnetic charges}

Considering the coordinates presented in~\cite{Podolsky:2021zwr,Podolsky:2022xxd} and the subsequent correction of the gauge field introduced in~\cite{Astorino:2023elf}, we start by writing down the PD metric as
\begin{eqnarray}
    \Omega^2\diff s^2_{0} &=& -\frac{Q}{R^2}\bb{\diff t - \bp{1-x}\bp{a+2l+ax}\diff\varphi}^2 + \frac{R^2}{1-x^2}\bp{\frac{\diff x^2}{P}+\frac{\bp{1-x^2}\diff r^2}{Q}}\nonumber\\
    &&+\frac{\bp{1-x^2}P}{R^2}\bbr{a\,\diff t - \bb{r^2+\bp{a+l}^2}\diff \varphi}^2,\label{eq:PDmetric}
\end{eqnarray}
where
\begin{subequations}\label{eq:PDmetric1}
\begin{align}
\Omega(r,x) & =1-\frac{aA}{a^2+l^2}\,r\bp{l+a x} \\
R^2(r,x) & =r^2+\bp{l+a x}^2, \\
P(x) & =\Omega(r_+,x)\,\Omega(r_-,x), \\
Q(r) & =\bp{r-r_{+}}\bp{r-r_{-}}\bp{1+aA\,\frac{a-l}{a^2+l^2}\,r}\bp{1-aA\, \frac{a+l}{a^2+l^2}\,r},
\end{align}
\end{subequations}
with
\begin{align}
r_{\pm}=m\pm\sqrt{m^2+l^2-a^2-e^2-g^2},
\end{align}
denoting the locations of the black hole horizons. The parameters appearing above are the ``physical'' ones: the mass $m$, {the Kerr-like rotation parameter $a$}, the acceleration parameter $A$, the NUT parameter $l$, and the electromagnetic charges $e$ and $g$.

We can cast~\eqref{eq:PDmetric} into the LWP form~\eqref{eq:e-LWPMain} with Weyl's canonical coordinates $\rho,z$ expressed in terms of $r,x$ via\footnote{For functions with a single argument, a prime accent denotes differentiation with respect to that argument.}
\begin{subequations}
    \begin{eqnarray}
    \rho(r,x)&=&\frac{\sqrt{\bp{1-x^2}PQ}}{\Omega^2},\\
        2A a^2 r^2 z(r,x)&=& \bp{a^2+l^2}\bb{r\bp{r_+ + r_-} -2 r_+ r_-}+2 aAlr_+ r_- r \nonumber\\
        &&+\frac{2\bb{a^2+l^2 - 2aA\bp{l+ a x}r}Q-\Omega\,Q'\bp{a^2+l^2}r}{ \Omega^2},
    \end{eqnarray}
and the seed functions being 
\begin{eqnarray}
    f_0(r,x)&=&\frac{Q-a^2\bp{1-x^2}P}{\Omega^2 R^2},\label{eq:f0Seed}\\
    \omega_0(r,x)&=&\bp{1-x}\bb{2l+a\bp{1+x}\bp{1-\frac{P}{\Omega^2f_0}}},\label{eq:om0Seed}\\
    \gamma(r,x)&=&\frac{1}{2}\ln\frac{R^2 f_0}{\Omega^2\bb{\bp{\partial_rz}^2+\bp{\partial_r\rho}^2}Q}.
\end{eqnarray}
\end{subequations}
Finally, the seed gauge field $A_0$ has nonvanishing temporal and azimuthal components,
\begin{subequations}
\begin{eqnarray}
A_{t,0}(r,x)&=&-\frac{er+g\bp{l+ax}}{R^2},\\
A_{\varphi,0}(r,x)&=&g x -\bp{1-x}\bp{a+2l+a x}A_{t,0},
\end{eqnarray}
\end{subequations}
respectively. 

This information almost suffices to identify the seed Ernst potentials. We also need to solve Eqs.~\eqref{eq:TwistEquations} for the seed twisted potentials $\tilde{A}_{\varphi,0}$ and $\chi_0$. To do so, we need to write the gradient of a function $F(r,x)$ in the coordinate system $\bbr{t,r,x,\varphi}$; it reads 
\begin{equation}
    \bn F = \frac{1}{h_r}\partial_r F \bv{r}+\frac{1}{h_x}\partial_x F \bv{x},
\end{equation}
where the scale factors $h_r(r,x)$ and $h_x(r,x)$ are given by
\begin{equation}
    h_r = \frac{R}{\Omega\sqrt{Q}}=h_x\sqrt{\frac{\bp{1-x^2}P}{Q}}.   
\end{equation}
Then, the differential equations via which we are to determine $\tilde{A}_{\varphi,0}$ become
\begin{subequations}
    \begin{eqnarray}
        \partial_x\tilde{A}_{\varphi,0}&=& a \,\partial_r A_{t,0},\\
        \partial_r\tilde{A}_{\varphi,0}&=& -\frac{\partial_x A_{t,0}}{a},
    \end{eqnarray}
\end{subequations}
admitting the solution 
\begin{equation}
    \tilde{A}_{\varphi,0}(r,x) = -\frac{gr - e\bp{l+ax}}{R^2},
\end{equation}
up, of course, to the addition of an integration constant, which we set to zero. The differential equations determining $\chi_0$, i.e., the second equation in the set~\eqref{eq:TwistEquations}, admit the solution 
\begin{eqnarray}
    \chi_0(r,x)&=&\frac{2\bp{a^2+l^2}\bb{amx -l\bp{r-m}}-2aA\bb{\bp{a^2-l^2}\bp{r-m}r+a\bp{2m-r_+}r_+\bp{l+ax}x}}{R^2\bp{a^2+l^2}\Omega},
\end{eqnarray}
again, up to the addition of an integration constant, which we also neglect for simplicity without loss of generality. The seed Ernst potentials are found by simply substituting the above functions into~\eqref{eq:ErnstPotentials}, and there is no reason to display them explicitly here.

Before proceeding with the Ehlers--Harrison transformation, let us communicate a somewhat interesting observation which will prove pertinent also in the target case. Notice that $\tilde{A}_{\varphi,0}$ is obtained from $A_{t,0}$ via a duality transformation of the charges $(e,g)\mapsto(g,-e)$. Indeed, the previous exchange of the charges generates a discrete phase transformation $\Phi_0\mapsto -i \Phi_0$ which maps $A_{t,0}\mapsto \tilde{A}_{\varphi,0}$ and $\tilde{A}_{\varphi,0}\mapsto - A_{t,0}$. Since $|\Phi_0|^2$ and $e^2+g^2$ are then invariant under such transformations, it follows that $\ernst_0$ is preserved. Therefore, looking at the target potentials~\eqref{eq:EHpotentials}, it becomes apparent that, if we simultaneously perform a duality transformation of the Harrison parameter $\beta=b_e+ib_m$, i.e., $(b_e,b_m)\mapsto(b_m,-b_e)$, or equivalently $\beta\mapsto -i\beta$, the target gravitational potential $\ernst$ is preserved whereas $\Phi\mapsto -i \Phi$, exactly as in the seed case. In other words, the particular exchanges of charges and parameters end up inducing a $\sfT^3_{-i}$ transformation of the potentials which, of course, leaves the Ernst equations invariant. Further looking at the twist equations~\eqref{eq:TwistEquations}, the differential equation determining $\omega$ does not transform, in contrast to the one determining the azimuthal component of the target gauge field. One can basically see where this is going; the target metric will be invariant under the simultaneous charge and Harrison-parameter duality exchanges, but the target Maxwell field will not, an after all quite expected result. 

Let us now operate on the seed potentials with a combined Ehlers--Harrison transformation via the composition $\sfE_c\circ\sfH_\beta$. The new potentials read 
    \begin{eqnarray}
        \ernst &=& \frac{\ernst_0}{\Lambda},\quad \Phi = \frac{\Phi_0 + \bp{b_e+i b_m}\ernst_0}{\Lambda},\\
        \Lambda &=& 1 + \bp{ic - b_e^2 - b_m^2}\ernst_0 - 2\bp{b_e - i b_m}\Phi_0,\quad \beta = b_e + i b_m\nonumber.
    \end{eqnarray}
Using the definitions~\eqref{eq:ErnstPotentials}, we can readily identify some of the new functions. In particular, 
\begin{subequations}
    \begin{eqnarray}
        f(r,x)&=&\frac{f_0}{|\Lambda|^2},\\
        |\Lambda|^2\chi(r,x)&=&\chi_0 - c |\ernst_0|^2 - 2b_e\bp{\chi_0A_{t,0}-\tilde{A}_{\varphi,0} \Re \ernst_0} -2 b_m \bp{\chi_0 \tilde{A}_{\varphi,0}+A_{t,0}\Re\ernst_0}, \\
        |\Lambda|^2A_t(r,x)&=& A_{t,0}+\bp{b_m c - b_e b_m^2-b_e^3}|\ernst_0|^2 +\bp{4 b_e b_m -c}\bp{\chi_0A_{t,0} - \tilde{A}_{\varphi,0}\Re\ernst_0} \nonumber\\
        &&-\bp{3 b_e^2 - b_m^2}\bp{\chi_0\tilde{A}_{\varphi,0} + A_{t,0}\Re\ernst_0}+b_e\bp{\Re\ernst_0 - 2 |\Phi_0|^2}-b_m\chi_0,\\
        \tilde{A}_\varphi(r,x)&=& A_{t}\mid \bbr{\Phi_0\mapsto -i\Phi_0,\beta\mapsto-i\beta},
    \end{eqnarray}
\end{subequations}
while $\gamma$ remains the same. We once again remind the reader that $b_e$ and $b_m$ are the real and imaginary parts of the Harrison parameter $\beta$, and $c$ is the real Ehlers parameter. Now, we still have to solve for $\omega$ and $A_{\varphi}$. {Ideally, we would like to obtain the result with all the seed charges and parameters switched on, but this turns out to be an extremely demanding task, computation-wise, in these coordinates. Therefore, we restrict ourselves to discussing a particular case whose novelty is sufficient in the sense that it corresponds to an accelerating and rotating black hole of Petrov type I with NUT parameter and electric charge, both coming from the Ehlers--Harrison map.

To this end, we switch off $b_m$, the seed NUT parameter $l$ and the initial charges $e,g$ in order to greatly simplify things.} With these assumptions, one can indeed first get $\omega$, 
    \begin{eqnarray}
        \omega &=&  \omega_0 +C_1 - \frac{a\bp{b_e^4+c^2}}{x^2}\bp{1-\frac{\bp{1-x^2}PF}{\Omega^5 R^2f_0}}-\frac{2c\bb{a^2+r\bp{r-2m}}H}{\Omega^4R^2f_0},
    \end{eqnarray}
where 
\begin{subequations}
\begin{eqnarray}
    F(r,x)&=& R^2\bbr{1-3Axrx-A^2\bb{r^2+a^2\bp{1+3x^2}} + A^3\bb{3r^2+a^2\bp{3+x^2}}xr }\nonumber\\
    &&-2m\bp{1-A^2 r^2}\bb{r\bp{1-x^2}+2x^2m -2A\bp{2r-m}xr - a^2A\bp{1+3x^2}x},\\
    H(r,x)&=&A\bp{1-x^2}R^2\bbr{1+A^2\bb{a^2\bp{1+x^2-2Axr}-r^2}}\nonumber\\
    &&-2m\bbr{1-2A xr +a^2A^2\bp{1-x^4} + 2A^3 \bb{r^2-a^2\bp{1-x^2}} xr -A^4 r^4}x,
\end{eqnarray}
\end{subequations}
are functions of $r,x$ which depend only on the seed parameters, and where $\omega_0$ is given in Eq.~\eqref{eq:om0Seed}. Note that the integration constant has been already shifted as to have a nonsingular $A\to 0$ limit. Having obtained $\omega$, we can now integrate the first equation in \eqref{eq:TwistEquations} for the azimuthal part of the target gauge field, finding 
\begin{equation}
    A_\varphi = G-\omega A_t + C_2,
\end{equation}
where
\begin{eqnarray}
    G&=&\frac{ab_e{}^3}{x^2}\left(1+\frac{2m\bp{1-x^2}P\bp{1-A^2r^2}\bb{r\bp{1-x^2}-4A r^2x - a^2A\bp{1+3x^2}x +2m\bp{x+rA}x}}{\Omega^5 R^2 f_0}\right. \nonumber\\
    &&-\left.\frac{\bp{1-x^2}P\bbr{1-3A rx-A^2\bb{r^2+a^2\bp{1+3x^2}}+A^3 x \bb{3r^2 + a^2\bp{3+x^2}} r}}{\Omega^5f_0}\right),
\end{eqnarray}
and $C_2$ is another integration constant.

Therefore, the target solution is given by
\begin{subequations}
    \begin{eqnarray}
        \diff s^2 &=& -\frac{f_0}{|1+\bp{ic-b_e^2}\ernst_0|^2}\bp{\diff t - \omega\, \diff \varphi}^2+\frac{|1+\bp{ic-b_e^2}\ernst_0|^2\bp{1-x^2}PQ}{\Omega^4f_0}\,\diff \varphi^2 \nonumber\\
        &&+\frac{|1+\bp{ic-b_e^2}\ernst_0|^2R^2}{\Omega^2Q}\bb{\diff r^2+\frac{Q}{\bp{1-x^2}P}\,\diff x^2},\label{eq:TheMetric}
    \end{eqnarray}
    and the gauge field 
    \begin{equation}
        A=-b_e\,\frac{b_e^2 |\ernst_0|^2 - f_0}{|1+\bp{ic-b_e{}^2}\ernst_0|^2}\,\diff t + \bp{G-\omega A_t+C_2}\diff \varphi.
    \end{equation}
\end{subequations}
With the restrictive assumptions $e=0=g=l$ we have made so far, we have
\begin{subequations}
    \begin{eqnarray}
        P(x)&=&1+A\bp{a^2Ax-2m}x,\\
        Q(r)&=&\bp{1-A^2 r^2}\bb{a^2+\bp{r-2m}r},\\
        \Omega(r,x)&=& 1- Axr,\\
        R(r,x)&=&\sqrt{r^2+a^2 x^2}.
    \end{eqnarray}
\end{subequations}
The seed function $f_0$ is given in~\eqref{eq:f0Seed}, whereas $\chi_0$ assumes the neat form 
\begin{equation}
    \chi_0(r,x) = \frac{2a\bb{m\bp{x+A r}-A R^2}}{\Omega R^2}.
\end{equation}
Thus, $\ernst_0 = f_0 + i \chi_0$, since the complex electromagnetic seed potential $\Phi_0$ is zero ($A_{t,0} =0 =A_{\varphi,0} = \tilde{A}_{\varphi,0}$).

Although the objective of providing explicit expressions is completed, the form~\eqref{eq:TheMetric} of the metric may not be the most convenient when discussing certain limits. Therefore, we also propose the alternative form
\begin{eqnarray}
        \Omega^2\diff s^2 &=& -\frac{Q - a^2 \Tilde{P}}{\mathcal{R}^2}\bb{\diff t - \bp{\tilde{\omega}+A\,\frac{\bp{1-x^2}\bp{r^2-2r\sqrt{\tilde{m}^2+\tilde{l}^2} +a^2} W}{4\bp{\tilde{m}^2+\tilde{l}^2}\bp{r^2-2r\sqrt{\tilde{m}^2+\tilde{l}^2}+a^2 x^2}\Omega^5 R^2 f_0}} \diff \varphi}^2\nonumber\\
        &&+\mathcal{R}^2\bp{\frac{\Tilde{P}Q}{Q-a^2\Tilde{P}}\,\diff \varphi^2 +\frac{\diff r^2}{Q}+\frac{\diff x^2}{\Tilde{P}}},\nonumber\\
        \label{eq:TheMetric2}
    \end{eqnarray}
where  
    \begin{eqnarray}
        \mathcal{R}^2(r,x) &=& \left|1-\bp{b_e^2+i\,\frac{\tilde{l}}{2\sqrt{\tilde{m}^2+\tilde{l}^2}}}\ernst_0\right|^2 R^2,\quad
        \tilde{P}(x) = \bp{1-x^2}P,\\
        \tilde{\omega}(r,x) &=& 2\tilde{l}\bp{1-x}-a\frac{8\tilde{m}^2\bp{\tilde{m}r+a\tilde lx}\bp{1-x^2}+\bp{\tilde l^2+4\tilde{m}^2b_e^4}\bbr{\bp{r-2\tilde{m}}^2+\bb{2\tilde{m}\bp{r-2\tilde{m}}+a^2}x^2}}{4\tilde{m}^2\bb{r\bp{r-2\tilde{m}}+a^2 x^2}}.\label{eq:DefsMetric2}
    \end{eqnarray}
with new parameters
\begin{equation}
    \Tilde{l}=-2cm,\quad \tilde{m} = m\sqrt{1-4c^2}.\label{eq:NewParams}
\end{equation}
Here, we have set $C_1 = 2\tilde{l}$ in order to avoid transforming the time coordinate which is the alternative course of action to obtain the above form if one wants to keep $C_1$ arbitrary. Moreover, the function $W(r,x)$ is quite involved. In particular, 
\begin{eqnarray}
    W(r,x)&=&-4m^2aR^2\Omega^3\bp{R^2A-2mx} +a\bp{\tilde{l}^2+4m^2 b_e^4}\tilde{F}+4m\tilde{l}\Omega \tilde{H},
\end{eqnarray}
with 
\bse{
    \tilde{F}(r,x)&=&8 m^3 \bp{x + r A} \bp{1 - r^2 A^2}+ 2R^2m \bbr{3 x + 3 r \bp{2 - x^2} A +  x \bb{2 a^2 \bp{1 + 3 x^2}-r^2 \bp{4 - x^2} } A^2} \nonumber\\
&&- 2 R^2 m \bbr{r \bb{3 r^2 \bp{2 - x^2} - a^2 \bp{1 - x^2 - 4 x^4}} A^3 + r^2 x \bb{ a^2 \bp{4 + x^2 -  x^4}-r^2 \bp{1 - x^2}} A^4} \nonumber \\ 
&& -  R^4 A \bp{3 - A \bbr{r x + \bb{3 r^2 - a^2 \bp{1 + 3 x^2}} A - r x \bb{r^2 -  a^2 \bp{3 + x^2}} A^2}} \nonumber \\ 
&& + 4 m^2 \bb{r^4 \bp{4 - 3 x^2} A^3 - a^2 x^2 \bp{2 + 3 x^2} A + r^5 x^3 A^4 + r^3 x A^2 \bp{4 -  x^2 + a^2 A^2}} \nonumber\\
&&+ 4 m^2 \bp{r^2 A \bbr{ x^2 \bb{3 + a^2 \bp{5 + x^2 -  x^4} A^2}-4}-r x \bb{4 + a^2 \bp{2 + 2 x^2 - 3 x^4} A^2}},\\
\tilde{H}(r,x)&=&2 R^2 m \bbr{- r -  x \bb{r^2 + a^2 \bp{2 + x^2}} A + r \bb{r^2 - a^2 \bp{1 - 3 x^2}} A^2 + r^2 x \bb{r^2 +  a^2 \bp{2 - x^2}} A^3} \nonumber \\ 
&& + 4 m^2 x \bp{r^3 A -  r^5 A^3 + a^2 \bbr{x + r A - r x A \bb{ x + r \bp{2 - x^2} A}}}\nonumber\\
&& -  R^4 \bbr{A^2 \bb{r^2 -  a^2 \bp{1 + x^2 - 2 r x A}}-1}.
}
{Observe that in some of the preceded equations, we used $m$ instead of $\tilde{m}$; this is not a typographical error. We rather did so only for brevity. Keep in mind that, whenever we express the metric in this form, the NUT parameter and the mass are given by $\tilde{l}$ and $\tilde{m}$ in Eq.~\eqref{eq:NewParams}, respectively, not by $c$ and $m$. This particular form of the target metric with the redefined parameters will be more suitable for various important limits, as we will see below.} 

In passing, we also remark that, unfortunately, the Misner string is not removable for $\tilde{l}\neq 0$, or $c\neq0$ in the original form~\eqref{eq:TheMetric}, since
\begin{equation}
    \Delta\omega = \lim\limits_{x\to 1}\omega - \lim\limits_{x\to -1}\omega = -4\tilde{l}.
\end{equation}
However, switching on the seed charges, we know that they will interact with the Harrison parameter $b_e$, contributing to the above discontinuity in such a way, that the latter becomes eliminable via proper tuning~\cite{Astorino:2023ifg}. Let us now start from this exotic enhanced Kerr metric of type I (a subfamily of the enhanced PD for $e=0=g=l$), and discuss some limiting cases.

\subsubsection*{Vanishing acceleration limit} 

In the case of vanishing acceleration, one expects to be able to recover the Kerr--Newman--NUT metric (type D). Indeed, after proper coordinate transformations and parameter redefinitions, the metric~\eqref{eq:TheMetric} acquires the form\footnote{Kerr--Newman--NUT metric as displayed in~\cite{Podolsky:2021zwr}.}
\begin{eqnarray}
    \diff s^2 &=& -\frac{Q}{R^2}\bb{\diff \bar{t} - \bp{1-x}\bp{\bar{a}+2\bar{l}+\bar{a}x}\diff\varphi}^2 + \frac{R^2}{1-x^2}\bp{{\diff x^2}+\frac{\bp{1-x^2}\diff \bar{r}^2}{Q}}\nonumber\\
    &&+\frac{\bp{1-x^2}}{R^2}\bbr{\bar{a}\,\diff \bar t - \bb{\bar{r}^2+\bp{\bar{a}+\bar{l}}^2}\diff \varphi}^2,
\end{eqnarray}
with
\begin{subequations}
    \begin{align}
R^2(r,x) & =\bar{r}^2+\bp{\bar l+\bar a x}^2, \\
Q(r) & =\bp{\bar r-\bar r_{+}}\bp{\bar r-\bar r_{-}},\\
\bar r_{\pm} &=\bar m \pm\sqrt{\bar m^2+\bar l^2-\bar a^2-\bar e^2},
\end{align}
\end{subequations}
where 
\begin{subequations}
    \begin{eqnarray}
        \bar{t}&=&\frac{1}{\sqrt{\bp{1-b_e^2}^2+c^2}}\bbr{t+\bb{a\bp{b_e^2+c^2}-4 cm - C_1}\varphi},\\
        \bar{r}&=&r\sqrt{\bp{1-b_e^2}^2+c^2} - \frac{2m\bb{c^2-b_e^2\bp{1-b_e^2}}}{\sqrt{\bp{1-b_e^2}^2+c^2}},\\
        \bar{a}&=&a \sqrt{\bp{1-b_e^2}^2+c^2},\\
        \bar{l}&=& -\frac{2cm}{\sqrt{\bp{1-b_e^2}^2+c^2}},\\
        \bar{m}&=& \frac{m\bp{1-c^2-b_e^4}}{\sqrt{\bp{1-b_e^2}^2+c^2}},\\
        \bar{e}&=& 2 m b_e.
    \end{eqnarray}
\end{subequations}
Clearly, the new NUT parameter and the new seed electric charge are proportional to the transformation parameters $c$ and $b_e$, respectively.\footnote{Do not confuse $\bar{m}$ with the $\Tilde{m}$ we previously introduced.} There is no need to display the gauge field here, for it will actually be misaligned with respect to the standard form in the Kerr--Newman--NUT solution. However, this issue is known, and its resolution is given by acting with an additional duality rotation on the Ernst electromagnetic potential~\cite{Astorino:2019ljy}.

\subsubsection*{Vanishing rotation limit}
The case of vanishing rotation corresponds to the enhanced C-metric (type I), i.e., a C-metric into which, NUT and electromagnetic charges enter via the Ehlers--Harrison map. Here, we choose to consider the alternative form~\eqref{eq:TheMetric2} of the target metric which is the most befitting for the task at hand. Therefore, our parameters are $\tilde{l},\tilde{m},a,A,b_e$. When $a\to 0$, the metric~\eqref{eq:TheMetric2} becomes 
\begin{eqnarray}
    \Omega^2\diff s^2 &=& -\frac{Q }{\mathcal{R}^2}\bbr{\diff t - \tilde{l}\bb{2\bp{1-x}+A\,\frac{\tilde{P}r^2}{\Omega^2 \sqrt{\tilde{m}^2 + \tilde{l}^2}} } \diff \varphi}^2+\mathcal{R}^2\bp{{\Tilde{P}}\,\diff \varphi^2 +\frac{\diff r^2}{Q}+\frac{\diff x^2}{\Tilde{P}}},
        \label{eq:TheMetric3}
\end{eqnarray}
where 
\begin{subequations}
\begin{equation}
    Q(r)=r\bp{r-2\sqrt{\tilde{m}^2 + \tilde{l}^2}}\bp{1-A^2r^2},\quad \tilde{P}(x)=\bp{1-x^2}\bp{1-2Ax\sqrt{\tilde{m}^2 + \tilde{l}^2}},\quad \Omega(r,x)=1-Axr,
\end{equation}
and 
\begin{equation}
    \mathcal{R}^2(r,x)=\frac{\Tilde{l}^2 Q^2 +4 \bp{\tilde{m}^2 + \tilde{l}^2} \bp{r^2\Omega^2 - b_e^2 Q}^2}{4 \bp{\tilde{m}^2 + \tilde{l}^2} \Omega^4 r^2}.
\end{equation}
\end{subequations}
The solution further contains a gauge field
\begin{eqnarray}
    A&=&b_e\,\frac{\bp{r^2\Omega^2 - b_e^2 Q}Q}{\mathcal{R}^2\Omega^4 r^2}\,\diff t + \bbr{C_2 - \bb{2\tilde{l}\bp{1-x}+A\,\tilde{l}\,\frac{\tilde{P}r^2}{\Omega^2 \sqrt{\tilde{m}^2 + \tilde{l}^2}}}A_t}\diff \varphi.
\end{eqnarray}
The explicit form~\eqref{eq:TheMetric3} of the enhanced C-metric is particularly suitable for taking further limits. In fact, it is not hard to observe that by killing the Harrison parameter $b_e$, the spacetime configuration assumes the form of the accelerating NUT black hole described in~\cite{Barrientos:2023tqb,Astorino:2023elf}. On the other hand, killing $\tilde{l}$, the solution reduces to the accelerating charged black holes described in \cite{Astorino:2023ifg}.
For pedagogical clarity, we explicitly highlight the limit pertaining to the accelerating-NUT spacetime discussed in reference \cite{Barrientos:2023tqb}. It is pertinent to note that the limit governing the spacetime outlined in reference \cite{Astorino:2023ifg} follows a similar logic. It proves convenient to start with the line element 
\begin{equation}
\Omega^2\diff s^2 = -\frac{Q }{\mathcal{R}^2}\bbr{\diff t +2c\bb{2m\bp{1-x}+A\,\frac{\tilde{P}r^2}{\Omega^2 } } \diff \varphi}^2+\mathcal{R}^2\bp{{\Tilde{P}}\,\diff \varphi^2 +\frac{\diff r^2}{Q}+\frac{\diff x^2}{\Tilde{P}}},
\end{equation}
where
\begin{equation}
\mathcal{R}^2=r^2+c^2\frac{Q^2}{\Omega^4r^2}, 
\end{equation}
and that simply descents from \eqref{eq:TheMetric3} by reverting the reparametrizations \eqref{eq:NewParams}. Defining 
\begin{equation}
\Tilde{m}=\sqrt{m^2-\bar{l}^2}
,\quad r_\pm=\tilde{m}\pm\sqrt{\tilde{m}^2+\bar{l}^2}, \quad c=\frac{\bar{l}}{r_+},
\end{equation} 
and performing the coordinates transformations
\begin{equation}
    t=\frac{r_+-r_-}{r_+}(\tilde{\tau}-2\bar{l}\varphi), \quad r=\tilde{r}-r_-,
\end{equation}
the spacetime metric converts into 
\begin{equation}
\frac{r_+}{r_+-r_-}\tilde{\Omega}^2\diff s^2=-\frac{\tilde{Q}}{\tilde{\mathcal{R}}^2}\left[\diff\tilde{\tau} -2\bar{l}\left(x-A\frac{\tilde{P}(\tilde{r}-r_-)^2}{(r_+-r_-)\tilde{\Omega}^2}\right)\diff\varphi \right]^2+\tilde{\mathcal{R}}^2\bp{{\tilde{P}}\,\diff \varphi^2 +\frac{\diff \tilde{r}^2}{\tilde{Q}}+\frac{\diff x^2}{\tilde{P}}},
\end{equation}
where the metric polynomials has been defined via 
\begin{equation}
\begin{aligned}
\tilde{\Omega}&=1-A(\tilde{r}-r_-)x,\\
\tilde{P}&=1-A(r_+-r_-)x,\\
\tilde{Q}&=(1-A^2(\tilde{r}-r_-)^2)(\tilde{r}-r_+)(\tilde{r}-r_-),\\
\tilde{\mathcal{R}}^2&=\frac{1}{r_+^2+\bar{l}^2}\left(r_+^2(\tilde{r}-r_-)^2+\bar{l}^2\frac{(\tilde{r}-r_+)^2(1-A^2(\tilde{r}-r_-)^2)^2}{\Omega^4}\right).
\end{aligned}
\end{equation}
Up to a standard rescaling of the whole metric by a constant conformal factor, this line element correctly reproduces the desired limit to the accelerating-NUT black hole \cite{Barrientos:2023tqb}. After implementing the aforementioned reparametrizations, it is essential to recall that the limit of vanishing mass leads us to the line element characterizing the massless accelerating-NUT black hole, and not, as naively expected, to the Ehlers Rindler spacetime. 

\subsubsection*{Transforming the Rindler spacetime}
Looking at the original form of the metric, Eq.~\eqref{eq:TheMetric}, with the involved parameters being $m,a,A,c,b_e$, we can proceed by directly killing the mass $m$ and the angular momentum $a$, this before any further reparametrizations, to obtain the Enhanced Rindler metric
\begin{equation}
\Omega^2\diff s^2 = -\frac{Q }{\mathcal{R}^2}\bb{\diff t + \frac{2cAr^2}{\Omega^2}\bp{1-x^2} \diff \varphi}^2+\mathcal{R}^2\bb{\bp{1-x^2}\diff \varphi^2 +\frac{\diff r^2}{Q}+\frac{\diff x^2}{1-x^2}},\label{eq:TheMetric4}
\end{equation}
where 
\begin{equation}
    Q(r)=r^2(1-A^2r^2),\quad \Omega(r,x)=1-Axr,\quad \mathcal{R}^2(r,x)=\frac{c^2 Q^2 +\bp{r^2\Omega^2 - b_e^2 Q}^2}{\Omega^4 r^2}.
\end{equation}
The gauge field accompanying it, reads 
\begin{eqnarray}
    A&=&b_e\frac{\bp{r^2\Omega^2 - b_e^2 Q}Q}{\mathcal{R}^2\Omega^4 r^2}\,\diff t + \bbr{C_2 +\frac{2cAr^2}{\Omega^2}\bp{1-x^2}A_t}\diff \varphi.
\end{eqnarray}
Interestingly, this spacetime is obtainable by essentially operating with a combined Ehlers--Harrison map on a Rindler spacetime. Therefore, it is evident that, if we switch off the acceleration, we recover Minkowski spacetime. Indeed, after a Weyl rescaling of the metric, $\diff s^2\to \diff \bar{s}^2 = \diff s^2/\bp{1-b_e^2 + c^2}$, and a time rescaling $t\to \bp{1-b_e^2 + c^2} \bar{t}$, this utterly proves to be the case. Of course, the Maxwell field also vanishes up to the choice of gauge. Similar to black holes in the previous sections, it can be demonstrated that the accelerating horizons are characterized by the incorporation of the Ehlers and Harrison parameters. This is achieved through appropriate reparametrizations and changes of coordinates. This outcome appears to be an inherent characteristic of these solutions, stemming from their conceptualization as accelerating black holes derived from a non-accelerating black hole binary wherein one of the black holes undergoes indefinite growth.

\section{Further comments}\label{seccoments} 

The present study endeavors to contribute to the discussion surrounding algebraically general black holes within the framework of Einstein--Maxwell theory. These type I spacetimes have recently gained substantial attention, mainly because they arise via a highly nontrivial action of the Ehlers or Harrison transformations on spacetimes featuring accelerating horizons. Specifically, operating on an accelerating seed with an Ehlers map, or a Harrison map, or both, has the remarkable effect of altering the algebraic properties of the seed, this due to the transformation parameters---a NUT parameter in the case of Ehlers, or electromagnetic charges in the case of Harrison---penetrating the Rindler horizon.

In this work, we presented a complete hierarchical structure for the type I solutions arising via the combined action of Ehlers and Harrison maps. The graphical form of the hierarchy was given in Fig.~\ref{FIG1}, with the graph's root node being the Enhanced Pleban\'ski--Demia\'nski spacetime, or EPD for short. We managed to provide an explicit form of the solution, at least up to the case of neutral seeds, a task computationally feasible only at the (minor) cost of using the original PD coordinates. However, in the case of a neutral NUTless PD seed, viz., the accelerating Kerr, we were able to integrate the equations directly in the physical spherical-like coordinates, thereby obtaining the explicit form of a novel type I spacetime representing accelerating and rotating black holes, endowed with both NUT and electromagnetic charges via the Ehlers--Harrison map. On top of that, we further scrutinized some limits of this solution, demonstrating how various spacetimes, previously presented in the pertinent literature, arise as limiting cases.

Before bringing forth the hierarchy of these type I solution, we engaged in a detailed investigation of the Ernst symmetries, particularly focusing on the Ehlers and Harrison maps. After reviewing how these two maps emerge from proper compositions of gravitational and electromagnetic gauge transformations with the inversion symmetry, inherent in the Ernst equations, we discussed their composition properties. It turned out that, although the Ehlers transformations form a subgroup, the Harrison ones do not, with the reason behind this failure made manifest. These insights into the transformations, as mathematical operations per se, provided a better understanding also of the physical effects they produce when acting on spacetimes. Finally, we thoughtfully included a user-friendly rederivation of the Ernst equations in Appendix~\ref{App2}, purely for pedagogical purposes, and in order to deal with minor inconsistencies, often encountered in the literature, regarding signs in the definitions of the Ernst potentials and the twisted-potential equations.

In considering ways to further enrich this type I hierarchy of solutions, an interesting prospect is the introduction of angular momentum into a given seed through a suitable solution-generating technique, as conjectured in~\cite{Astorino:2023ifg}. A promising approach would involve the inverse scattering method~\cite{Belinski:2001ph}, a mechanism known for generating, e.g. the Kerr spacetime from the Minkowski metric. If the Rindler horizon somehow interacts with the external angular momentum, one can speculate that the entire type I hierarchy, as presented here, would be further extended to allow for two distinct angular momenta, the seed one and the one introduced via the solution-generating technique. Consequently, the intriguing thought of a Rindler--Kerr background may ultimately materialize, \emph{inter alia} becoming a fertile soil for exploring novel interactions between all parameters within this generalized family.

Although searching for ways to extend this hierarchy is definitely tempting, it is readily evident that there are numerous novel geometries within the EPD family, which need to be thoroughly examined. A comprehensive investigation of their causal structure, along with a satisfactory geometric description of how they extend beyond their type D counterparts, is imperative. In addition, given the presence of acceleration, delving into their thermodynamics constitutes an intriguing challenge. A succinct framework for understanding the thermodynamics of accelerating black holes, remains yet elusive, recent commendable contributions towards this direction \cite{Ball:2021xwt, Anabalon:2018ydc, Anabalon:2018qfv, Astorino:2016ybm, Kim:2023ncn} notwithstanding.\footnote{Recently the thermodynamics of accelerating black holes in three-dimensions has been also explored, opening a new road towards exploring the holographic properties of accelerating spacetimes~\cite{Arenas-Henriquez:2022www,Arenas-Henriquez:2023vqp,Arenas-Henriquez:2023hur,Cisterna:2023qhh}.} Last but not least, it is necessary to fully probe and understand the intricate mechanism behind the change in the algebraic nature of an accelerating seed, effected by Ehlers or Harrison transformations acting on the latter---namely how these operations alter the principal null directions of the Weyl tensor, etc.---, in an attempt to solidify a consistent framework for generating algebraically general black holes.

\acknowledgments 

The work of  J.B.  is supported by FONDECYT Postdoctorado grant No. 3230596.  A.C. is partially supported by FONDECYT grant 1210500 and by PRIMUS/23/SCI/005 and GA{\v C}R 22-14791S grants from Charles University. K.P. acknowledges financial support provided by the European Regional Development Fund (ERDF) through the Center of Excellence TK133 ``The Dark Side of the Universe'' and PRG356 ``Gauge gravity: unification, extensions and phenomenology''.

\appendix 

\section{EPD spacetime in the original PD coordinates: the neutral-seed case}\label{App1}

In this section, we present a detailed construction of the Enhanced Pleban\'ski--Demia\'nski spacetime. To facilitate this task, two key assumptions are made. First, we employ the original PD coordinates to lighten the computational burden. {Second, in order to obtain an analytic non-integral form of the target twisted potentials, we shall switch off the seed electromagnetic charges, $e$ and $g$. Despite these simplifying assumptions, our setup is yet general enough to accommodate, for the first time, an explicit integration of the solution obtained by acting on a rotating seed with the Ehlers--Harrison map.} 

Let us start by writting down the line element of the PD spacetime, and the gauge field supporting it, in the following way,
\begin{subequations}
\begin{align}
    \diff s{}^2_0 &= -f_0\bp{\diff t-\omega_0 \,\diff\varphi}^2 + \frac{\rho^2\,\diff \varphi^2}{f_0}+ \frac{R^2}{\Omega^2}\bp{\frac{\diff r^2}{Q} + \frac{\diff x^2}{P}},\label{eq:TheMetricPDcoords}\\
    A_0&=-\frac{er + \hat{\omega}gx}{R^2}\,\diff t+ \frac{e \hat{\omega} r x^2 - gxr^2}{R^2}\,\diff \varphi,
\end{align}
\end{subequations}
where we have introduced the functions 
\begin{subequations}
    \begin{eqnarray}
        f_0(r,x) &=& \frac{Q-\hat{\omega}^2P}{\Omega^2 R^2},\\
        \omega_0(r,x) &=& \oo\frac{x^2Q+r^2P}{Q-\oo P},\\
        \rho(r,x) &=& \frac{\sqrt{PQ}}{\Omega^2},\\
        Q(r)&=&\oo^2 k + e^2 + g^2 - 2mr +\epsilon r^2 - \frac{2 A n r^3}{\oo}-kA^2r^4,\\
        P(x)&=&k+\frac{2nx}{\oo}-\epsilon x^2 +2 A m x^3 - A^2\bp{k\oo^2+e^2+g^2}x^4,\\
        R(r,x)&=&\sqrt{r^2+\oo^2x^2},\\
        \Omega(r,x)&=&1-A xr.
    \end{eqnarray}
\end{subequations}
Note that the above form of the PD family depends on seven parameters $(m,n,e,g, A,\hat{\omega},k)$. It is often
assumed that $m$, $n$, $e$, and $g$ represent the mass, NUT, electric and
magnetic charges, respectively. {The parameter $\hat{\omega}$, often called the twist parameter, plays a crucial role. It is associated with both angular momentum and NUT charge \cite{Griffiths:2005se,Griffiths:2005mi,Griffiths:2005qp}.
A reconstruction of the same spacetime was performed later by Podolsk\'y and Vr\'atn\'y \cite{Podolsky:2021zwr,Podolsky:2022xxd}, where they introduced a refined coordinate system where the twist parameter $\hat{\omega}$ can be effectively absorbed, simplifying the analysis of these spacetimes.  Specifically, it was determined in \cite{Vratny:2023bst}, and subsequently utilized in \cite{Podolsky:2021zwr,Podolsky:2022xxd}, that the most convenient choice for the twist parameter is $\hat{\omega}=(a^2+l^2)/a^2$. By adopting this particular value for $\hat{\omega}$ we arrive at the PD metric as described by equations  \eqref{eq:PDmetric} and \eqref{eq:PDmetric1} which involves only six parameters  $(m,l,e,g,A,a)$.}
 
As it stands, the metric~\eqref{eq:TheMetricPDcoords} is already in the 
Lewis--Papapetrou form~\eqref{eq:e-LWPMain}, albeit in the chart $\{t,r,x,\varphi\}$, with Weyl's coordinate $\rho$ given above, and the $z$ coordinate given by
\begin{equation}
    z(r,x)=\frac{k\oo A r^2+n\bp{1+Axr}r-\oo A \bb{A\bp{e^2+g^2+k\oo^2}x-m\bp{1+Axr}+\epsilon r}x}{\oo\Omega^2}.
\end{equation}
For completeness, we also display the function $\gamma$, 
\begin{equation}
    \gamma(r,x)=\frac{1}{2}\ln \frac{Q-\oo^2P}{\Omega^4\bb{P\bp{\partial_x\rho}^2+Q\bp{\partial_r\rho}^2}}.
\end{equation}
To find the target metric we first need to solve Eqs.~\eqref{eq:TwistEquations} for the twisted potentials which read
\bse{
    \tilde{A}_{\varphi,0}&=&\frac{gr-e\oo x}{R^2},\\
    \chi_0&=&2\frac{nr - m\oo x + \oo A\bb{kr^2 + \bp{e^2+g^2+k\oo^2}x^2}}{\Omega R^2}.
}
With these at hand, the target metric functions, i.e., the ones obtained after the application of the Ehlers--Harrison map, can be expressed as follows
\bse{
        f&=&\frac{f_0}{|\Lambda|^2},\\
       \oo^2|\Lambda|^2R^2\Omega^4\chi&=& -2 \hat{\omega}^2 (-1 + r x A)^3 \bigl [n r -  m \hat{\omega} x + k \hat{\omega} (r^2 + \hat{\omega}^2 x^2) A\bigl ]\nonumber\\
    &&- c \biggl(4 n^2 (\hat{\omega}^2 + r^4 A^2) + 4 m^2 (\hat{\omega}^2 + \hat{\omega}^4 x^4 A^2) - 4 n \hat{\omega} \bigl \{- k A \bigl [- \hat{\omega}^4 x^3 A + r^5 A^2 \nonumber\\
    &&\phantom{- c \biggl(}+ \hat{\omega}^2 r (2 - 3 r x A + r^2 x^2 A^2)\bigl ] + (\hat{\omega}^2 x + r^3 A) \epsilon \bigl \} - 4 m \hat{\omega} \bigl \{-2 n (r^2 + \hat{\omega}^2 x^2) A\nonumber \\
    &&\phantom{- c \biggl(}+ k \hat{\omega} A \bigl [- r^3 A + \hat{\omega}^4 x^5 A^2 + \hat{\omega}^2 x (2 - 3 r x A + r^2 x^2 A^2)\bigl ] + \hat{\omega} (r + \hat{\omega}^2 x^3 A) \epsilon \bigl \}\nonumber \\
    &&\phantom{- c \biggl(}+ \hat{\omega}^2 (r^2 + \hat{\omega}^2 x^2) \bigl \{k^2 A^2 \bigl [r^4 A^2 + \hat{\omega}^4 x^4 A^2 + 2 \hat{\omega}^2 (2 - 4 r x A + r^2 x^2 A^2)\bigl ] \nonumber\\
    &&\phantom{- c \biggl(}- 2 k (r^2 -  \hat{\omega}^2 x^2) A^2 \epsilon + \epsilon^2\bigl \}\biggr),\\
    \bp{\oo^2/b_e}|\Lambda|^2R^2\Omega^4 A_t&=&\hat{\omega} (1-A r x)^2 \bigl \{-2 n (\hat{\omega}^2 x + r^3 A) - 2 m \hat{\omega} (r + \hat{\omega}^2 x^3 A)\bigl\}  \nonumber\\
    &&+ \hat{\omega}^2(1-Axr)^2 (r^2 + \hat{\omega}^2 x^2) \bigl [k (- r^2 + \hat{\omega}^2 x^2) A^2 + \epsilon \bigl ]\nonumber\\
    &&-  b_e^2 \biggl(4 n^2 (\hat{\omega}^2 + r^4 A^2) + 4 m^2 (\hat{\omega}^2 + \hat{\omega}^4 x^4 A^2) - 4 n \hat{\omega} \bigl \{- k A \bigl [- \hat{\omega}^4 x^3 A + r^5 A^2 \nonumber\\
    &&\phantom{-b_e^2 \biggl(}+ \hat{\omega}^2 r (2 - 3 r x A + r^2 x^2 A^2)\bigl ] + (\hat{\omega}^2 x + r^3 A) \epsilon \bigl \} - 4 m \hat{\omega} \bigl \{-2 n (r^2 + \hat{\omega}^2 x^2) A \nonumber\\
    &&\phantom{-b_e^2 \biggl(}+ k \hat{\omega} A \bigl [- r^3 A + \hat{\omega}^4 x^5 A^2 + \hat{\omega}^2 x (2 - 3 r x A + r^2 x^2 A^2)\bigl ] + \hat{\omega} (r + \hat{\omega}^2 x^3 A) \epsilon \bigl \}\nonumber \\
    &&\phantom{-b_e^2 \biggl(}+ \hat{\omega}^2 (r^2 + \hat{\omega}^2 x^2) \bigl \{k^2 A^2 \bigl [r^4 A^2 + \hat{\omega}^4 x^4 A^2 + 2 \hat{\omega}^2 (2 - 4 r x A + r^2 x^2 A^2)\bigl ]\nonumber\\
    &&\phantom{-b_e^2 \biggl(}- 2 k (r^2 -  \hat{\omega}^2 x^2) A^2 \epsilon + \epsilon^2\bigl \}\biggr),\\
   \tilde{A}_{\varphi}&=& b_e \chi.
 }

The rotational function $\omega$ can be divided in two terms. The first term is the seed function $\omega_0$, which stands for the rotational function of the Kerr--NUT black hole. The second term contains all the couplings with the Ehlers and Harrison parameters. Thus,
\begin{equation}
    \omega=\omega_0+\varpi+C_1,
\end{equation}
where $\varpi$ reads  
\begin{eqnarray}
    \varpi&=& -  \frac{2 c (k \hat{\omega} + 2 n x) -  b_e^4 k^2 \hat{\omega}^2 A -  c^2 k^2 \hat{\omega}^2 A}{\hat{\omega} x^2 A} \nonumber\\
    &&-  \frac{\bigl \{k \hat{\omega} (-1 + \hat{\omega}^2 x^4 A^2) + x \bigl [-2 n + \hat{\omega} x (-2 m x A + \epsilon)\bigl ]\bigl \} }{\hat{\omega} x^2 A (-1 + r x A)^3 \bigl \{-2 n (\hat{\omega}^2 x + r^3 A) - 2 m \hat{\omega} (r + \hat{\omega}^2 x^3 A) + \hat{\omega} (r^2 + \hat{\omega}^2 x^2) \bigl [k (- r^2 + \hat{\omega}^2 x^2) A^2 + \epsilon \bigl ]\bigl \}}\nonumber\\
    &&\times \biggl(2 c (-1 + r x A) \bigl \{2 m \hat{\omega} r (-1 + 2 r x A + \hat{\omega}^2 x^4 A^2) + 2 n \bigl [r^3 A (-1 + 2 r x A) + \hat{\omega}^2 x (-1 + r x A + r^2 x^2 A^2)\bigl ] \nonumber\\
    &&\phantom{\times \biggl(}+ \hat{\omega} (r^2 + \hat{\omega}^2 x^2) \bigl [k A^2 (- r^2 -  \hat{\omega}^2 x^2 + 2 r^3 x A) + \epsilon - 2 r x A \epsilon \bigl ]\bigl \} + b_e^4 A \bigl \{k^2 \hat{\omega}^2 (r^2 + \hat{\omega}^2 x^2)\nonumber\\
    &&\phantom{\times \biggl(}\times A^2 (- r^2 - 3 \hat{\omega}^2 x^2+ 3 r^3 x A + \hat{\omega}^2 r x^3 A) + k \hat{\omega} \bigl [2 n r^3 A (-1 + 4 r x A) + 2 n \hat{\omega}^2 x (-1 -  r x A + 3 r^2 x^2 A^2)\nonumber\\
    &&\phantom{\times \biggl(}+ 2 m \hat{\omega} (- r + 3 r^2 x A + 3 \hat{\omega}^2 x^3 A -  r^3 x^2 A^2) -  \hat{\omega} (r^2 + \hat{\omega}^2 x^2) (-1 + 3 r x A) \epsilon \bigl ]\nonumber\\
    &&\phantom{\times \biggl(}+ 2 x (n r -  m \hat{\omega} x) \bigl [2 m \hat{\omega} + r (2 n r A -  \hat{\omega} \epsilon)\bigl ]\bigl \} + c^2 A \bigl \{k^2 \hat{\omega}^2 (r^2 + \hat{\omega}^2 x^2) A^2 (- r^2 - 3 \hat{\omega}^2 x^2 
    + 3 r^3 x A + \hat{\omega}^2 r x^3 A) \nonumber\\
    &&\phantom{\times \biggl(} k \hat{\omega} \bigl [2 n r^3 A (-1 + 4 r x A) + 2 n \hat{\omega}^2 x (-1 -  r x A + 3 r^2 x^2 A^2) + 2 m \hat{\omega} (- r + 3 r^2 x A + 3 \hat{\omega}^2 x^3 A -  r^3 x^2 A^2) \nonumber\\
    &&\phantom{\times \biggl(}-  \hat{\omega} (r^2 + \hat{\omega}^2 x^2) (-1 + 3 r x A) \epsilon \bigl ]+ 2 x (n r -  m \hat{\omega} x) \bigl [2 m \hat{\omega} + r (2 n r A -  \hat{\omega} \epsilon)\bigl ]\bigl \}\biggr).
\end{eqnarray}
Finally, it remains to integrate the magnetic component of the target gauge field, which is found to be 
\begin{equation}
    A_\varphi =G-\omega A_t + C_2,
\end{equation} 
with 
\begin{eqnarray}
    G(r,x) &=&  \frac{ b_e^3 \bigl \{k \hat{\omega} (-1 + \hat{\omega}^2 x^4 A^2) + x \bigl [-2 n + \hat{\omega} x (-2 m x A + \epsilon)\bigl ]\bigl \}}{\hat{\omega} x^2 (-1 + r x A)^3 \bigl \{-2 n (\hat{\omega}^2 x + r^3 A) - 2 m \hat{\omega} (r + \hat{\omega}^2 x^3 A) + \hat{\omega} (r^2 + \hat{\omega}^2 x^2) \bigl [k (- r^2 + \hat{\omega}^2 x^2) A^2 + \epsilon \bigl ]\bigl \}}\nonumber\\
    &&\times \biggl(k^2 \hat{\omega}^2 (r^2 + \hat{\omega}^2 x^2) A^2 (- r^2 - 3 \hat{\omega}^2 x^2 + 3 r^3 x A + \hat{\omega}^2 r x^3 A) + k \hat{\omega} \bigl [2 n r^3 A (-1 + 4 r x A)\nonumber \\
    &&\phantom{\times \biggl(}+ 2 n \hat{\omega}^2 x (-1 -  r x A + 3 r^2 x^2 A^2) + 2 m \hat{\omega} (- r + 3 r^2 x A + 3 \hat{\omega}^2 x^3 A -  r^3 x^2 A^2) \nonumber\\
    &&\phantom{\times \biggl(}-  \hat{\omega} (r^2 + \hat{\omega}^2 x^2) (-1 + 3 r x A) \epsilon \bigl ]+ 2 x (n r -  m \hat{\omega} x) \bigl [2 m \hat{\omega} + r (2 n r A -  \hat{\omega} \epsilon)\bigl ]\biggr) - \frac{b_e^3 k^2 \hat{\omega}}{x^2},
\end{eqnarray}
and $C_2$ being yet another integration constant.

\section{A user-friendly guide to the Ernst formalism}\label{App2}

In this section, we present a detailed derivation of the renowned Ernst equations. It all starts with the Einstein-Maxwell action\footnote{We use natural units, and we further have set $G=(4\pi)^{-1}$.}
\begin{equation}
    I_{\mathrm{EM}}\bb{g_{\mu\nu},A_\mu}=\frac{1}{4}\int\diff^4x\sqrt{-g}\bp{R-F_{\mu\nu}F^{\mu\nu}}.
\end{equation}
Varying with respect to the metric and the gauge field we obtain the field equations 
\begin{subequations}
    \begin{eqnarray}
        G_{\mu\nu}&=&2\bp{F_{\mu}{}^\lambda F_{\nu\lambda} - \frac{1}{4}F_{\lambda\sigma}F^{\lambda\sigma} g_{\mu\nu}},\\
        \partial_\nu\bp{\sqrt{-g} F^{\nu\mu}}&=&0,
    \end{eqnarray}
\end{subequations}
respectively. Since the trace of the energy-momentum tensor vanishes, the metric field equations admit a particularly simple Ricci form, 
\begin{equation}
    R_{\mu\nu}=2\bp{F_{\mu}{}^\lambda F_{\nu\lambda} - \frac{1}{4}F_{\lambda\sigma}F^{\lambda\sigma} g_{\mu\nu}}=:2 T_{\mu\nu}.
\end{equation}

\subsection{Field equations with the ``electric" LWP ansatz}
Now, since we are interested in stationary and axisymmetric spacetimes characterized by two commuting
Killing vectors, $\partial_t$ and $\partial_\varphi$, we consider the LWP metric ansatz
\begin{equation}
    \diff s^2 = -f\bp{\diff t-\omega \,\diff\varphi}^2 + \frac{1}{f}\bb{\rho^2\diff\varphi^2 + \expo{2\gamma}\bp{\diff \rho^2 + \diff z^2}},\label{eq:e-LWP}
\end{equation}
together with a gauge field with the same symmetries
\begin{equation}
    A= A_t\,\diff t + A_\varphi\,\diff\varphi,\label{eq:MaxwellAnsatz}
\end{equation}
where $f,\,\omega,\,\gamma$ and $A_t,\,A_\varphi$ are functions of $\rho$ and $z$.

The simplest equations to tackle first are the field equations for the Maxwell field $A_\mu$. One can easily show that 
\begin{equation}
    F_{\mu\nu}=\delta^{\rho t}_{\mu\nu}A'_t+\delta^{zt}_{\mu\nu}\dot{A}_t+\delta^{\rho\varphi}_{\mu\nu}A'_\varphi+\delta^{z\varphi}_{\mu\nu}\dot{A}_\varphi,
\end{equation}
where our convention for the rank-4 skew-symmetric Kronecker delta is $\delta^{\lambda\sigma}_{\mu\nu}=\delta^\lambda_\mu\delta^\sigma_\nu - \delta^\lambda_\nu\delta^\sigma_\mu$. A prime accent $^\prime$ denotes a derivative with respect to $\rho$, and a dot accent $\dot{}$ denotes a derivative with respect to $z$. Since $\partial_t,\,\partial_\varphi$ are Killing vectors, we have that only the vectors $F^{\rho\mu}$ and $F^{z\mu}$ appear in the Maxwell field equations. These read
\begin{subequations}
    \begin{eqnarray}
        \sqrt{-g} F^{\rho\mu} &=& \rho\bb{-\frac{A'_t}{f}+\frac{\omega f}{\rho^2}\bp{A'_\varphi + \omega A'_t}}\delta^\mu_t+{f\rho}\frac{A'_\varphi+\omega A'_t}{\rho^2}\delta^\mu_\varphi=:\rho \hat{F}^\mu,\\
        \sqrt{-g} F^{z\mu} &=& \rho\bb{-\frac{\dot{A}_t}{f}+\frac{\omega f}{\rho^2}\bp{\dot{A}_\varphi + \omega \dot{A}_t}}\delta^\mu_t+{f\rho}\frac{\dot{A}_\varphi+\omega \dot{A}_t}{\rho^2}\delta^\mu_\varphi=:\rho F^\mu.
    \end{eqnarray}
\end{subequations}
Then, the Maxwell field equations are given by 
\begin{equation}
    \frac{1}{\rho}\bp{\rho \hat{F}^\mu}'+ \dot{F}^\mu=0.
\end{equation}
Given that the divergence of a vector $\mathbf{V}(\rho,z)$ in cylindrical coordinates reads  
\begin{equation}
\boldsymbol{\nabla}\cdot \mathbf{V}=\frac{1}{\rho}(\rho V_\rho)'+\dot{V}_z,
\end{equation}
it turns out the Maxwell field equations can be written as 
\begin{equation}\label{defFmu}
\boldsymbol{\nabla}\cdot \mathbf{F}^\mu=0,
\end{equation}
where 
\begin{equation}
\mathbf{F}^\mu:=\bp{\hat{F}^\mu,F^\mu,0}=\bb{-\frac{\boldsymbol{\nabla}A_t}{f}+\frac{\omega f}{\rho^2}\bp{\boldsymbol{\nabla}A_\varphi + \omega \boldsymbol{\nabla}A_t}}\delta^\mu_t+{f}\frac{\boldsymbol{\nabla}A_\varphi+\omega \boldsymbol{\nabla}A_t}{\rho^2}\delta^\mu_\varphi.
\end{equation}
Let us put now our attention on Einstein field equations. In particular, we have the $tt$ component
\begin{eqnarray}\label{eqEEtt} -2f^3\bp{\boldsymbol{\nabla} A_\varphi +\omega \boldsymbol{\nabla}A_t}\cdot\bp{\boldsymbol{\nabla} A_\varphi +\omega \boldsymbol{\nabla}A_t}-\rho^2 \boldsymbol{\nabla}f\cdot\boldsymbol{\nabla}f+f^4\boldsymbol{\nabla}\omega\cdot\boldsymbol{\nabla}\omega
    -f\rho^2\bp{2\boldsymbol{\nabla}A_t\cdot \boldsymbol{\nabla}A_t-\nabla^2 f}=0,\label{eq:Einstein1}
\end{eqnarray}
and the $t\varphi$ component
\begin{eqnarray}\label{eqEEtvarphi}
    -f^4\omega \nqform{\omega}+2 \omega f^3 \qform{\bp{\boldsymbol{\nabla} A_\varphi +\omega \boldsymbol{\nabla}A_t}}&\nonumber\\
    -f\rho^2\bb{\omega\bp{2\nqform{A_t}+\nabla^2f}+2\bp{2\boldsymbol{\nabla}A_\varphi \cdot \boldsymbol{\nabla}A_t +\boldsymbol{\nabla}f\cdot \boldsymbol{\nabla}\omega}}
    -f^2\rho\bp{\rho\nabla^2\omega -2\omega'}+\omega\rho^2\nqform{f}&=0.
\end{eqnarray}
Looking at the form of these two, it becomes apparent that, multiplying Eq. \eqref{eqEEtt} by $\omega$ and adding to Eq. \eqref{eqEEtvarphi}, we can obtain a simpler equation, namely, 
\begin{eqnarray}
    \frac{2f}{\rho^2}\bp{2\omega \nqform{A_t} + 2\boldsymbol{\nabla}A_\varphi\cdot\boldsymbol{\nabla}A_t+\boldsymbol{\nabla}f\cdot \boldsymbol{\nabla}\omega}+\frac{f^2}{\rho^2}\bp{\nabla^2\omega -\frac{2\omega'}{\rho}}=0.
\end{eqnarray}
At this stage, and after  cumbersome algebra, we can finally present the latter as 
\begin{equation}
    \boldsymbol{\nabla}\cdot\bb{\frac{f^2}{\rho^2}\boldsymbol{\nabla}\omega + \frac{4f}{\rho^2}A_t\bp{\boldsymbol{\nabla}A_\varphi + \omega \boldsymbol{\nabla}A_t}}=0,\label{eq:Einstein2}
\end{equation}
modulo the Maxwell field equations. Knowing all the other functions, we can also determine $\gamma$ via the equations
\begin{equation}
    r\bp{R_{\rho z}-2T_{\rho z}}=0,\quad \frac{r}{2}\bb{R_{\rho\rho}-R_{zz}-2\bp{T_{\rho\rho}-T_{zz}}}=0,\label{eq:GammaQuadratures}
\end{equation}
which directly provide us with expressions for $\dot{\gamma}$ and $\gamma'$, respectively. Do also note that if the $tt$ and $t\varphi$ components of the Einstein field equations are satisfied, then the $\varphi\varphi$ component vanishes identically.

\subsection{Let's twist again}

Notice that if $h$ is some function of $\rho,z$, then 
\begin{equation}
    \boldsymbol{\nabla}\cdot \bp{\frac{1}{\rho}\hat{\boldsymbol{\varphi}}\times \boldsymbol{\nabla}h}=0,\label{eq:Id1}
\end{equation}
regardless of how our triad is ordered. Therefore, in a fashion similar to ``closed is locally exact'', here we may argue that $\boldsymbol{\nabla}\cdot \mathbf{V}(\rho,z)=0$ implies that there exists a function $h(\rho,z)$ such that
\begin{equation}
    \mathbf{V}=\frac{1}{\rho}\hat{\boldsymbol{\varphi}}\times \boldsymbol{\nabla}h.
\end{equation}
Such a function will be called a \emph{twisted potential}. Now, recalling the definition of our vectors $\mathbf{F}^t$ and $\mathbf{F}^\varphi$ \eqref{defFmu}, and considering one of the Maxwell field equations, namely $\boldsymbol{\nabla}\cdot \mathbf{F}^\varphi=0$, we can always write
\begin{equation}
    \rho \mathbf{F}^\varphi = (-)^p\hat{\boldsymbol{\varphi}}\times \boldsymbol{\nabla}\Tilde{A}_\varphi.\label{eq:TwistedEM}
\end{equation}
Here, $p=0,\,1$ is introduced just to keep track of the available sign freedom in the definition of the twisted potential. Since $\hat{\boldsymbol{\varphi}}\times \bp{\hat{\boldsymbol{\varphi}}\times \mathbf{V}}=-\mathbf{V}$ for any vector $\mathbf{V}$ with only $\rho,z$ components, we may cross both sides of the above equation with $\hat{\boldsymbol{\varphi}}$ from the left to get 
\begin{equation}
    \frac{(-)^p}{f}\boldsymbol{\nabla}\Tilde{A}_\varphi+\frac{\omega}{\rho} \bv{\varphi}\times \boldsymbol{\nabla}A_t=-\frac{1}{\rho}\bv{\varphi}\times \boldsymbol{\nabla} A_\varphi.
\end{equation}
Using Eq.~\eqref{eq:Id1}, we can easily show that 
\begin{equation}
    \boldsymbol{\nabla}\cdot \bp{\frac{(-)^p}{f}\boldsymbol{\nabla}\Tilde{A}_\varphi+\frac{\omega}{\rho} \bv{\varphi}\times \boldsymbol{\nabla}A_t}=0,\label{eq:Max1Alternative}
\end{equation}
which can freely replace the equation $\boldsymbol{\nabla}\cdot\mathbf{F}^\varphi=0$. Remember that the other equation in the Maxwell set reads
\begin{equation}
    \boldsymbol{\nabla}\cdot \bp{-\frac{1}{f}\boldsymbol{\nabla}A_t+\omega \mathbf{F}^\varphi}=0.
\end{equation}
Clearly, using the definition of the twisted potential, we can present it as 
\begin{equation}
    \boldsymbol{\nabla}\cdot \bp{-\frac{1}{f}\boldsymbol{\nabla}A_t+(-)^p\frac{\omega}{\rho} \bv{\varphi}\times \boldsymbol{\nabla} \Tilde{A}_\varphi}=0.\label{eq:Max2Alternative}
\end{equation}
Hence, we have managed to cast the Maxwell field equations $\boldsymbol{\nabla}\cdot \mathbf{F}^\mu=0$ into a pair of equations comprised of~\eqref{eq:Max1Alternative} and~\eqref{eq:Max2Alternative}. Multiplying Eq.~\eqref{eq:Max1Alternative} with $i$ and subtracting Eq.~\eqref{eq:Max2Alternative} from it, we can express our pair as the single complex equation 
\begin{equation}
    \bn\cdot\bp{\frac{1}{f}\bn\Phi + \frac{i\omega}{\rho}\bv{\varphi}\times\bn\Phi}=0,\label{eq:MaxwellComplex}
\end{equation}
where $\Phi=A_t+i(-)^p\Tilde{A}_\varphi$ is the first complex potential we have introduced. 

Let us now turn back our attention towards the gravity sector. Using the identity \eqref{eq:Id1}, we can show that 
\begin{equation}
    (-)^p\bn \cdot\bp{\frac{1}{\rho}\bv{\varphi}\times \Tilde{A}_\varphi \bn A_t}=-\frac{1}{2}\bn\cdot\bb{\frac{1}{\rho}\bv{\varphi}\times \Im\bp{\Phi^\ast \bn \Phi}}.\label{eq:Id2}
\end{equation}
Taking into account the above equation, then Eq. \eqref{eq:Einstein2} can  be written as 
\begin{equation}
\bn\cdot\bp{\frac{f^2}{\rho^2}\bn\omega +4A_t\mathbf{F}^{\varphi}}=0.
\end{equation}
Using the definition~\eqref{eq:TwistedEM} together with the identity~\eqref{eq:Id2}, and considering that 
\begin{equation}
\Im\bp{\Phi^\ast \bn\Phi}=(-)^p\bp{A_t\bn\Tilde{A}_\varphi - \Tilde{A}_\varphi\bn A_t},
\end{equation}
one can cast Eq. \eqref{eq:Einstein2} into 
\begin{equation}
\bn\cdot\bb{\frac{f^2}{\rho^2}\bn\omega + \frac{2}{\rho}\bv{\varphi}\times\Im\bp{\Phi^*\bn\Phi}}=0.
\end{equation}
Following the same procedure as for Maxwell's equations, we introduce another twisted potential $\chi$ such that
\begin{equation}
    \frac{f^2}{\rho}\bn\omega + 2\bv{\varphi}\times\Im\bp{\Phi^*\bn\Phi}=(-)^s\bv{\varphi}\times\bn\chi,\label{eq:TwistedGrav}
\end{equation}
where $s=0,\,1$ is another parameter introduced to keep track of the sign freedom in the definition of the second twisted potential. If we  cross now both sides of the above equation with $\hat{\boldsymbol{\varphi}}$ from the left, then  
\begin{equation}
    \frac{1}{f^2}\bb{(-)^s\bn\chi -2 \Im\bp{\Phi^\ast \bn\Phi}}=-\frac{1}{\rho}\bv{\varphi}\times \bn\omega.
\end{equation}
Again, using the identity~\eqref{eq:Id1}, one can easily show that the equation
\begin{equation}
    \bn\cdot \left\lbrace \frac{1}{f^2}\bb{(-)^s\bn\chi -2 \Im\bp{\Phi^\ast \bn\Phi}}\right\rbrace=0,\label{eq:Einstein2Alt}
\end{equation}
may freely replace the Einstein equation~\eqref{eq:Einstein2}. For later use, we denote the above as $\bn\cdot \mathbf{G}=0$.

At this stage, we need to recall that for a vector $\mathbf{V}$ with only $\rho,z$ components, it holds that 
\begin{equation}
    \bp{\bv{\varphi}\times \mathbf{V}}\cdot\bp{\bv{\varphi}\times \mathbf{V}}=\qform{\mathbf{V}}.
\end{equation}
We may look at the other Einstein equation, namely Eq.~\eqref{eq:Einstein1}, and use this knowledge together with the definition of $\mathbf{F}^{\varphi}$, to write it as 
\begin{eqnarray}\label{eqA12rewrittten}
    -2\rho^4 f \qform{\mathbf{F}^\varphi}-\rho^2\nqform{f}+f\rho^2\nabla^2 f-2f\rho^2\nqform{A_t}+\rho^2 f^4 \qform{\mathbf{G}}&=&0.
\end{eqnarray}
Considering that 
\begin{equation}
\rho^2\qform{\mathbf{F}^{\varphi}}=\nqform{\Tilde{A}_\varphi},\qquad \bn\Phi\cdot\bn\Phi^* = \nqform{A_t}+\nqform{\Tilde{A}_\varphi},
\end{equation}
then Eq. \eqref{eqA12rewrittten} becomes 
\begin{eqnarray}
    -\nqform{f}+f\nabla^2 f-2f\bn\Phi\cdot\bn\Phi^*+ f^4 \qform{\mathbf{G}}&=&0.\label{eq:Einstein1Alt}
\end{eqnarray}
Multiplying Eq.~\eqref{eq:Einstein2Alt} with $if^2$ and subtracting it from Eq.~\eqref{eq:Einstein1Alt}, we combine the two Einstein equations into one complex gravitational equation, namely 
\begin{equation}
    -\nqform{f}+f\nabla^2 f-2f\bn\Phi\cdot\bn\Phi^*+ f^4 \qform{\mathbf{G}}-if^2\bn\cdot\mathbf{G}=0.\label{eq:EinsteinComplex}
\end{equation}
\subsection{The Ernst equations}
Recall now that the defining equation for the twisted potential $\chi$, Eq.~\eqref{eq:TwistedGrav}, can be written as 
\begin{equation}
    \frac{1}{\rho}\bn\omega = \bv{\varphi}\times \mathbf{G},\label{eq:omegaG}
\end{equation}
and by using the identity \eqref{eq:Id1}, the Maxwell equations \eqref{eq:MaxwellComplex} can be brought to the form 
\begin{equation}
    -\frac{1}{f^2}\bn f\cdot \bn \Phi + \frac{1}{f}\nabla^2\Phi + \frac{i}{\rho}\bn\omega\cdot\bp{\bv{\varphi}\times \bn\Phi}=0.
\end{equation}
Remembering the product property
\begin{equation}
\mathbf{X}\cdot\bp{\mathbf{Y}\times \mathbf{Z}}=-\mathbf{Z}\cdot\bp{\mathbf{Y}\times \mathbf{X}},
\end{equation}
which holds true for arbitrary vectors $\mathbf{X}$, $\mathbf{Y}$, and $\mathbf{Z}$, and using Eq. \eqref{eq:omegaG}, we finally reach 
\begin{equation}
     f\nabla^2\Phi =\bn f\cdot \bn \Phi -i f^2\bn\Phi\cdot\mathbf{G}.\label{eq:MaxwellComplexAlt}
\end{equation}

At this stage, if we make an educated introduction of a complex gravitational potential 
\begin{equation}
    \ernst = f-|\Phi|^2-i(-)^s\chi,
\end{equation}
we can show that Eq. \eqref{eq:MaxwellComplexAlt} can be written as 
\begin{equation}
    \bp{\Re\ernst + |\Phi|^2}\nabla^2\Phi = \bn\Phi\cdot\bp{\bn\ernst + 2\Phi^*\bn\Phi}.\label{eq:Ernst2}
\end{equation}
Most interesting, however, is the fact that after the introduction of the potential $\ernst$, the complex gravitational equation \eqref{eq:EinsteinComplex} also takes a similar form, namely 
\begin{equation}
    \bp{\Re\ernst + |\Phi|^2}\nabla^2\ernst = \bn\ernst\cdot\bp{\bn\ernst + 2\Phi^*\bn\Phi},\label{eq:Ernst1}
\end{equation}
---{modulo} Eq. \eqref{eq:Ernst2}. The remaining pair of Einstein equations, which gives $\gamma$ in terms of integrals, can also be expressed in terms of the two complex potentials. The form of these equations will not bother us here since we are going to determine $\gamma$ in a different manner via comparison.  

Summing up the findings, we introduced two complex potentials, 
\begin{equation}
    \ernst = f-|\Phi|^2-i(-)^s\chi,\quad \Phi = A_t+i(-)^p\Tilde{A}_\varphi,
\end{equation}
and two twisted potentials $\Tilde{A}_\varphi$ and $\chi$, which are given by the equations
\begin{subequations}
\begin{eqnarray}
\hat{\boldsymbol{\varphi}}\times \boldsymbol{\nabla}\Tilde{A}_\varphi &=& \frac{(-)^p f}{\rho}\bp{\bn A_\varphi + \omega \bn A_t},\\
    \bv{\varphi}\times\bn\chi &=&(-)^s\bp{\frac{f^2}{\rho}\bn\omega + 2\bv{\varphi}\times\Im\bp{\Phi^*\bn\Phi}},
\end{eqnarray}
\end{subequations}
respectively. With these at hand, we showed that the Einstein--Maxwell system assumes the form of a pair of complex equations, known as the Ernst equations \cite{Ernst:1967wx,Ernst:1967by}, 
\begin{subequations}\label{eq:ErnstEquations}
    \begin{eqnarray}
        \bp{\Re\ernst + |\Phi|^2}\nabla^2\ernst &=& \bn\ernst\cdot\bp{\bn\ernst + 2\Phi^*\bn\Phi},\\
        \bp{\Re\ernst + |\Phi|^2}\nabla^2\Phi &=& \bn\Phi\cdot\bp{\bn\ernst + 2\Phi^*\bn\Phi}.
    \end{eqnarray}
\end{subequations}

\subsection{Making the ``magnetic'' LWP ansatz}
After a double Wick rotation, the metric~\eqref{eq:e-LWP} acquires the so-called ``magnetic'' form 
\begin{equation}
    \diff s^2 = f\bp{\diff \varphi-\omega \diff t}^2 + \frac{1}{f}\bb{\expo{2\gamma}\bp{\diff \rho^2 + \diff z^2}-\rho^2\diff t^2 }.\label{eq:m-LWP}
\end{equation}
Considering a Maxwell field of the form \eqref{eq:MaxwellAnsatz}, it is possible to show that the Maxwell equations read $\bn\cdot\mathbf{F}^\mu=0$ again, with the difference being in the definition of $\mathbf{F}^\mu$. In particular, now we have
\begin{equation}
    \mathbf{F}^\mu={f}\frac{\boldsymbol{\nabla}A_t+\omega \boldsymbol{\nabla}A_\varphi}{\rho^2}\delta^\mu_t+\bb{-\frac{\boldsymbol{\nabla}A_\varphi}{f}+\omega \mathbf{F}^t}\delta^\mu_\varphi.
\end{equation}
Following the method used in the ``electric'' case, we now define a twisted potential $\Tilde{A}_t$ via the equation
\begin{equation}
    \rho \mathbf{F}^t=(-)^p\bv{\varphi}\times \Tilde{A}_t.
\end{equation}
Clearly, equation
\begin{equation}
    \boldsymbol{\nabla}\cdot \bp{\frac{(-)^p}{f}\boldsymbol{\nabla}\Tilde{A}_t+\frac{\omega}{\rho} \bv{\varphi}\times \boldsymbol{\nabla}A_\varphi}=0,
\end{equation}
is now the alternative form of $\bn\cdot \mathbf{F}^t=0$. If we define our complex potential as 
\begin{equation}\label{eqA51}
    \Phi=A_\varphi + i(-)^p\Tilde{A}_t,
\end{equation}
we can show that the Maxwell field equations assume the form of the complex equation~\eqref{eq:MaxwellComplex}.

If we now move our attention to the Einstein equations, we have that  the $\varphi\varphi$ component  can be written as
\begin{eqnarray}\label{eqA52}
     -2f^3\qform{\bp{\bn A_t + \omega \bn A_\varphi}}+\rho^2\nqform{f}-f^4\nqform{\omega}
     -f\rho^2\bp{2\nqform{A_\varphi}+\nabla^2 f}=0,
\end{eqnarray}
whereas the $t\varphi$ component assumes the form
\begin{eqnarray}\label{eqA53}
    f^4\omega \nqform{\omega}+2 \omega f^3 \qform{\bp{\boldsymbol{\nabla} A_t +\omega \boldsymbol{\nabla}A_\varphi}}&\nonumber\\
    -f\rho^2\bb{\omega\bp{2\nqform{A_\varphi}-\nabla^2f}+2\bp{2\boldsymbol{\nabla}A_\varphi \cdot \boldsymbol{\nabla}A_t -\boldsymbol{\nabla}f\cdot \boldsymbol{\nabla}\omega}}
    +f^2\rho\bp{\rho\nabla^2\omega -2\omega'}-\omega\rho^2\nqform{f}&=0.
\end{eqnarray}
Multiplying Eq. \eqref{eqA52} by $\omega$ and adding it to Eq. \eqref{eqA53}, we obtain
\begin{eqnarray}
   \frac{2f}{\rho^2}\bp{\boldsymbol{\nabla}f\cdot \boldsymbol{\nabla}\omega -2\omega \nqform{A_\varphi} - 2\boldsymbol{\nabla}A_\varphi\cdot\boldsymbol{\nabla}A_t}+\frac{f^2}{\rho^2}\bp{\nabla^2\omega -\frac{2\omega'}{\rho}}=0.
\end{eqnarray}
Modulo the Maxwell field equations, this takes the neat form 
\begin{equation}
   \boldsymbol{\nabla}\cdot\bb{\frac{f^2}{\rho^2}\boldsymbol{\nabla}\omega - \frac{4f}{\rho^2}A_\varphi\bp{\boldsymbol{\nabla}A_t + \omega \boldsymbol{\nabla}A_\varphi}}=0.\label{eq:mag-Einstein2}
\end{equation}
The $tt$ component vanishes identically if the $\varphi\varphi$ and $t\varphi$ components are satisfied, and once again, $\gamma$ is given in terms of integrals by solving Eqs.~\eqref{eq:GammaQuadratures}.

Using the new definition of the complex potential $\Phi$, Eq. \eqref{eqA51}, then the equation \eqref{eq:mag-Einstein2} can be easily cast into 
\begin{equation}
    \bn\cdot\bb{\frac{f^2}{\rho^2}\bn\omega - \frac{2}{\rho}\bv{\varphi}\times\Im\bp{\Phi^*\bn\Phi}}=0.
\end{equation}
Therefore, our twisted potential $\chi$ is given by 
\begin{equation}
    \frac{f^2}{\rho}\bn\omega - 2\bv{\varphi}\times\Im\bp{\Phi^*\bn\Phi}=(-)^s\bv{\varphi}\times\bn\chi,\label{eq:mag-TwistedGrav}
\end{equation}
and using the identity~\eqref{eq:Id1}, one can show that the equation
\begin{equation}
    \bn\cdot \left\lbrace \frac{1}{f^2}\bb{(-)^s\bn\chi +2 \Im\bp{\Phi^\ast \bn\Phi}}\right\rbrace=0,\label{eq:mag-Einstein2Alt}
\end{equation}
is now the one that may replace the Einstein equation~\eqref{eq:mag-Einstein2}. Again, we denote the above as $\bn\cdot \mathbf{G}=0$.

In the same fashion as in the electric case, we can express the $\varphi\varphi$ component of the Einstein field equations as 
\begin{eqnarray}
   -\nqform{f}+f\nabla^2 f+2f\bn\Phi\cdot\bn\Phi^*+ f^4 \qform{\mathbf{G}}=0.\label{eq:mag-Einstein1Alt}
\end{eqnarray}
We also see that the Maxwell field equations can be once again cast into the form~\eqref{eq:MaxwellComplexAlt}. Introducing now the complex gravitational potential 
\begin{equation}
    \ernst = -f-|\Phi|^2+i(-)^s\chi,
\end{equation}
we observe that the Einstein--Maxwell system can be brought to the form of the Ernst equations~\eqref{eq:ErnstEquations}. Summing up the findings in the magnetic case, we introduced two complex potentials, 
\begin{equation}
    \ernst = -f-|\Phi|^2+i(-)^s\chi,\quad \Phi = A_\varphi+i(-)^p\Tilde{A}_t,
\end{equation}
and two twisted potentials $\Tilde{A}_t$ and $\chi$, which are given by the equations
\begin{subequations}
\begin{eqnarray}
    \hat{\boldsymbol{\varphi}}\times \boldsymbol{\nabla}\Tilde{A}_t &=& \frac{(-)^p f}{\rho}\bp{\bn A_t + \omega \bn A_\varphi},\\
    \bv{\varphi}\times\bn\chi &=&(-)^s\bp{\frac{f^2}{\rho}\bn\omega - 2\bv{\varphi}\times\Im\bp{\Phi^*\bn\Phi}},
\end{eqnarray}
\end{subequations}
respectively.

\bibliography{apssamp}

\end{document}